\documentclass[twocolumn]{aastex63}
\usepackage{fontawesome}
\usepackage{CJK}
\usepackage{natbib}
\graphicspath{{figures}}

\newcommand{\teff}{T$_{\rm eff}$}
\newcommand{\feh}{\rm{[Fe/H]}}

\received{January 18, 2021}
\revised{May, June 2021}
\accepted{\today}

\submitjournal{AJ}

\shorttitle{Transiting Planets Around Halo Stars I}
\shortauthors{Kolecki et al.}

\graphicspath{{./}{figures/}}

\begin{document}
\begin{CJK*}{UTF8}{gbsn}
\title{Searching For Transiting Planets Around Halo Stars. I. Sample Selection and Validation}

\author{Jared R. Kolecki}\affiliation{Department of Astronomy, The Ohio State University, Columbus, Ohio 43210, USA}

\author[0000-0002-4361-8885]{Ji Wang (王吉)}\affiliation{Department of Astronomy, The Ohio State University, Columbus, Ohio 43210, USA}

\author{Jennifer A. Johnson}\affiliation{Department of Astronomy, The Ohio State University, Columbus, Ohio 43210, USA}

\author{Joel C. Zinn}\altaffiliation{NSF Astronomy and Astrophysics Postdoctoral Fellow.}\affiliation{Department of Astronomy, The Ohio State University, Columbus, Ohio 43210, USA}\affiliation{Department of Astrophysics, American Museum of Natural History, Central Park West at 79th Street, New York, NY 10024, USA}

\author{Ilya Ilyin}\affiliation{Leibniz-Institute for Astrophysics Potsdam (AIP), An der Sternwarte 16, D-14482 Potsdam, Germany}

\author{Klaus G. Strassmeier}\affiliation{Leibniz-Institute for Astrophysics Potsdam (AIP), An der Sternwarte 16, D-14482 Potsdam, Germany}

\begin{abstract}

By measuring the elemental abundances of a star, we can gain insight into the composition of its initial gas cloud---the formation site of the star and its planets. Planet formation requires metals, the availability of which is determined by the elemental abundance. In the case where metals are extremely deficient, planet formation can be stifled. To investigate such a scenario requires a large sample of metal-poor stars and a search for planets therein. This paper focuses on the selection and validation of a halo star sample. We select $\sim$17,000 metal-poor halo stars based on their Galactic kinematics, and confirm their low metallicities ($\feh < -0.5$), using spectroscopy from the literature. Furthermore, we perform high-resolution spectroscopic observations using LBT/PEPSI and conduct detailed metallicity ([Fe/H]) analyses on a sample of 13 previously known halo stars that also have hot kinematics. We can use the halo star sample presented here to measure the frequency of planets and to test planet formation in extremely metal-poor environments. The result of the planet search and its implications will be presented and discussed in a companion paper by Boley et al.   \\

\end{abstract}
\section{Introduction}

When and under what conditions did the first planet form? The oldest planetary system that we know of is $\sim$11.2 Gyr old~\citep[Kepler-444, ][]{Campante2015,Mack+2018}, with a host star metallicity of $\feh = -0.52$. The first generation of stars and planets is expected to have metallicities similar to or below that of Kepler-444. Such a metal-poor environment poses challenges for planet formation. For gas giant planets, core mass growth via accretion halts in the extremely metal-poor regime, resulting in a strong dependence of planet occurrence rate on stellar metallicity---known as the planet-metallicity correlation~\citep[e.g.,][]{Fischer2005}. For terrestrial planets, the planet-metallicity correlation is weaker~\citep{Wang2015} and period-dependent~\citep{Wilson2018,Petigura2018}. Therefore, metallicity is a key to regulate planet formation. More interestingly, some elements (e.g., $\alpha$-elements) may play a bigger role than others~\citep{Adibekyan2012,Bashi2019}. 

The abundance of the elements increases as the Milky Way evolves. Among the stellar populations in the Milky Way, halo stars are the most metal-poor on average and therefore serve as an ideal test ground for planet formation in metal-poor environments. However, identifying halo stars is difficult because halo stars are relatively rare in the solar neighborhood, representing of order 1\% of the local population; discovering them has historically been laborious; and detailed characterization via high-resolution spectroscopy has been time-consuming (see \citet{Carney+Latham} and associated works). 

The Gaia mission~\citep{Gaia2016} significantly changed the landscape of the study of halo stars, because we can use it to identify metal-poor stars based on their kinematics, thanks to the well-known correlation between the two \citep{els}. Gaia's astrometric and radial velocity measurements of one billion stars in the Milky Way provide an efficient way of identifying halo stars. As such, we set out to identify stars with halo kinematics using the Gaia DR2 data~\citep{Gaia2018}. The kinematically selected halo stars are then validated using data from the APOGEE spectroscopic survey~\citep{Majewski} and other photometric surveys to check the purity of the sample. 

Many of the halo stars are in the field of view of the Transiting Exoplanet Survey Satellite~\citep[TESS,][]{Ricker2015}. Searching for transiting planets around halo stars are therefore made possible by the Gaia and TESS missions. Our halo sample is much larger than previous samples used for planetary searches via radial velocity~\citep{Sozzetti2006, Faria2016}. Since $>70\%$ stars in our sample have [Fe/H]$<-$1, our search will likely result in planetary systems that are more metal-poor than the current record holder at [Fe/H]=$-0.84$~\citep{Escude2014}. The methods and results of the planet search are described in a companion paper (Boley et al. 2021). 

It is of interest to determine the absolute amount of metal deficiency of the protoplanetary disk as input into simulations of planet
formation in disks \citep[e.g.,][]{johnson2012}. However, halo stars in the literature can have values that differ by $\sim$ a factor
of two, even when all studies use high-resolution/high signal-to-noise data \citep[e.g.,][]{jofre2014}. Among the sources of systematic differences in the studies, the validity of using ionization equilibrium in LTE to determine gravity \citep{Thevenin1999} is another area where Gaia information is crucial. Gaia parallaxes resolve the question of whether a star is a subgiant or main-sequence star. We therefore consider a selection of well-studied halo stars from the literature with new spectra and analyses in this work.

The paper is organized as follows: Section \ref{Sample Selection} outlines the process and criteria used to select a sufficiently pure sample of halo dwarf stars, while Section \ref{Sample Validation} offers verification of the success of the sample selection. Section \ref{Observations} details the observation and data reduction process used to generate the spectra used in Section \ref{Abundance Analysis} to analyze the iron abundances of selected halo stars from the literature. The results of this analysis are discussed and compared with other papers in Section \ref{Discussion}.

\section{Sample Selection} \label{Sample Selection}

\begin{deluxetable*}{|c c c c c c c|}
\centering
\tablehead{
\colhead{Gaia DR2 ID} &
\colhead{pmRA (mas/yr)} &
\colhead{pmDEC (mas/yr)} &
\colhead{$T_{mag}$} &
\colhead{$M/M_{\odot}$} &
\colhead{$L/L_{\odot}$} &
\colhead{[M/H] from TIC}
}

\startdata
688442476835805312 & 66.074 & -114.073 & 13.068 & 1.11 & 0.92 & -1.52 \\
800209689226496128 & -234.096 & -159.769 & 14.759 & 0.61 & 0.04 & -1.11 \\
1021587525024433152 & -46.409 & -55.106 & 14.499 & 1.26 & 0.59 & -1.82 \\
808143181016203264 & 60.511 & -128.934 & 14.341 & 1.03 & 0.63 & -1.43 \\
\enddata
\caption{Selected entries from this paper's final halo catalog. The entire catalog is available electronically as a csv file, featuring additional columns not present in this print version of the table.} \label{table:Catalog}
\end{deluxetable*}

Stars were selected preferentially from the CTLv08\textsubscript{01} \citep{stassun+2018,stassun+2019} to have dwarf-like radii (raddflag = 1); have Gaia DR2 parallaxes and proper motions (\texttt{PMFlag} == `gaia2'; \texttt{PARFlag} == `gaia2'; \texttt{gaiaqflag} == 1); be within 1kpc according to the TICv8, which uses the \citet{BailerJones2018} method to calculate distance \citep{stassun+2019} (this is to ensure that the kinematic cut described below is valid, which was designed based on a simulation of the nearest 1kpc stars); and be outside of the ecliptic plane ($|\texttt{lat}| > 6$) to ensure candidates are observable by TESS. Note that, although the majority of targets from the CTL have $T < 13$, the Cool Dwarf list extends to $T = 16$ \citep[for a description of this special list, see][]{muirhead+2018}.

Additional stars were selected by an equivalent search in the TICv8 \citep{stassun+2018,stassun+2019} for stars with $13 \leq T < 15$ (otherwise the same cuts as for the CTL are applied), effectively extending the magnitude limit of the halo dwarf search to $T < 15$. 
 
To select stars with halo-like kinematics, we ran a Milky Way \texttt{Galaxia} \citep{sharma+2011} simulation of stars within 1kpc. We designed a kinematic cut to choose halo stars based on only proper motion using this simulation, so that a Gaia radial velocity was not required. Our \texttt{Galaxia} simulations were run using the default parameters, which are described in \cite{sharma+2011}. In particular, the \texttt{Galaxia} stellar halo geometry follows the Besan\c{c}on model of an oblate spheroid, with density proportional to $\left(\frac{\mathrm{Max}(a_c,a)}{R_{\odot}}\right)^{n}$, where $a^2 = R^2 + \frac{z^2}{\epsilon^2}$, $a_c = 500\  \mathrm{pc}$,  $\epsilon=0.76$, and $n=-2.44$ \citep{robin+2003}, with $R$ and $z$ denoting Galactocentric cylindrical radius and height above the plane in pc. Following \cite{bond+2010}, the \texttt{Galaxia} halo velocity distribution is assumed to be an ellipsoid with $\sigma_R = 141\ \mathrm{km}\,\mathrm{s}^{-1}$ and $\sigma_{\theta} = \sigma_{\phi} = 75\  \mathrm{km}\,\mathrm{s}^{-1}$, which describes well the observed SDSS halo velocity distribution within 10kpc \citep{smith+2009} as well that of Gaia halo stars \citep{posti+2018}.

The resulting kinematic selection was applied to the CTL/TIC sample, such that $v_{\delta} < 0\ \mathrm{km}/\mathrm{s}$ and either $v_{\alpha} < -230\ \mathrm{km}/\mathrm{s}$, $v_{\alpha} > 210\ \mathrm{km}/\mathrm{s}$, or $v_{\delta} < -\sqrt{(1 - ((v_{\alpha} + 10)\times 220)^2)} \times 230\ \mathrm{km}/\mathrm{s}$.

\begin{figure}\centering
 \includegraphics[width=\columnwidth]{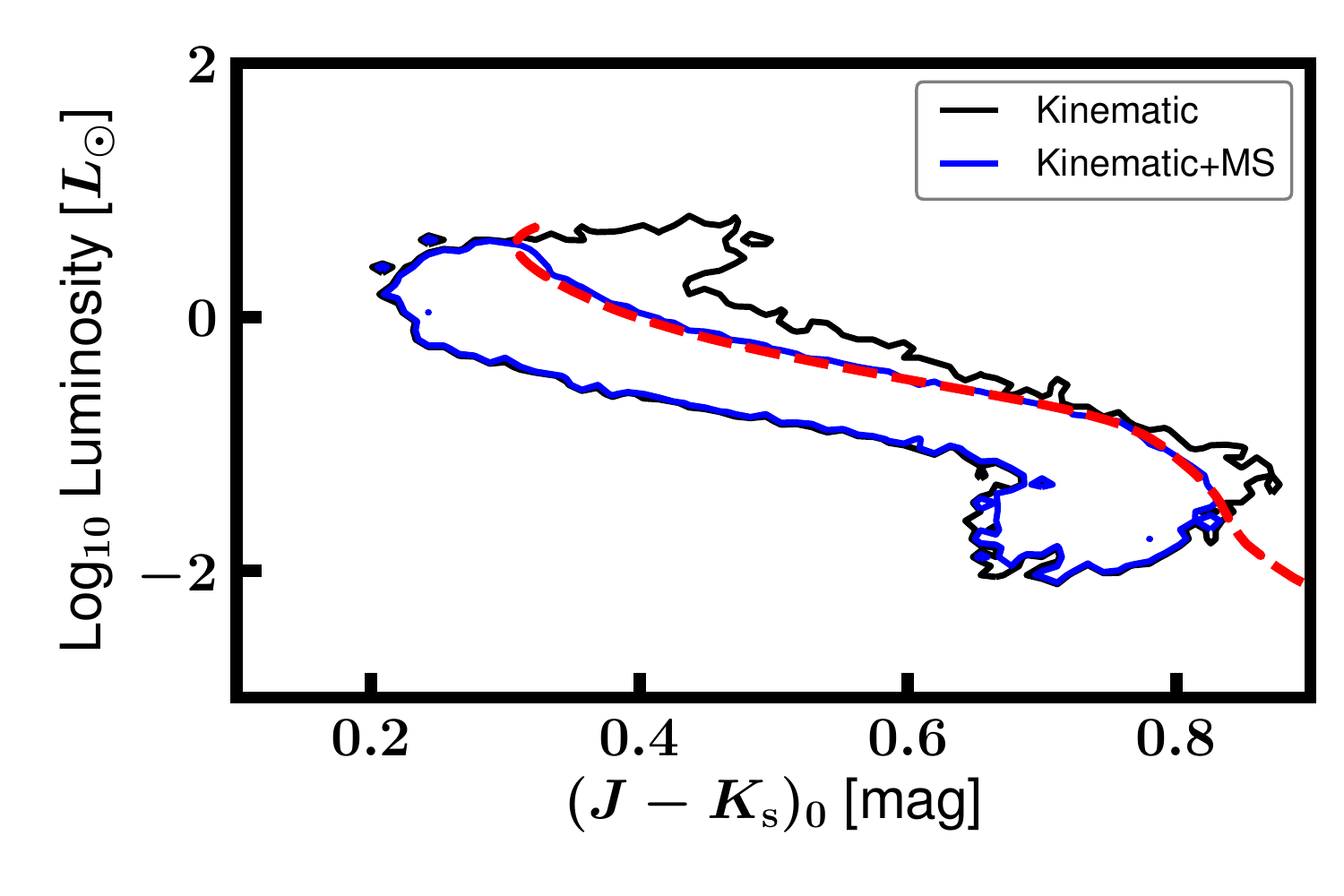}
 \caption{Color-luminosity diagram demonstrating our H-R Diagram selection criteria. Dwarf halo candidates passing kinematic selection criteria are shown as a black contour, representing the 95\% percentile. The binary main sequence of a $\feh = -1$ isochrone is shown as the dashed curve, and the distribution of the dwarf halo candidates falling below this sequence and therefore chosen as our final sample, is shown as the blue contour.}\label{fig:select}
\end{figure}

Finally, stars were selected to fall below the binary sequence of a MIST 13 Gyr $\feh = -1.0$ \citep{dotter+2016a,choi+2016a} isochrone in $J-K_{\rm{s}}$--$L$ space, where the luminosity, $L$, is adopted from the TIC. The final number of halo dwarf candidates selected in this way is 16940. The stars passing the kinematic cuts versus those passing the additional binary main sequence cut are shown in Figure~\ref{fig:select}. An excerpt from the final sample is displayed in Table \ref{table:Catalog}. The entire catalog is available electronically in csv format.

\section{Sample Validation}\label{Sample Validation}

Our sample selection can be tested by looking at the subset of stars with full 3-D kinematics or with metallicities. While the goal of the sample is to locate metal-poor stars, confirming that the proper-motion selected stars fall into kinematically hot populations when radial velocities are added is also valuable, given the correlation between the two properties.

\subsection{Galactic Velocity Distribution}
\begin{figure*}
\centering
 \includegraphics[]{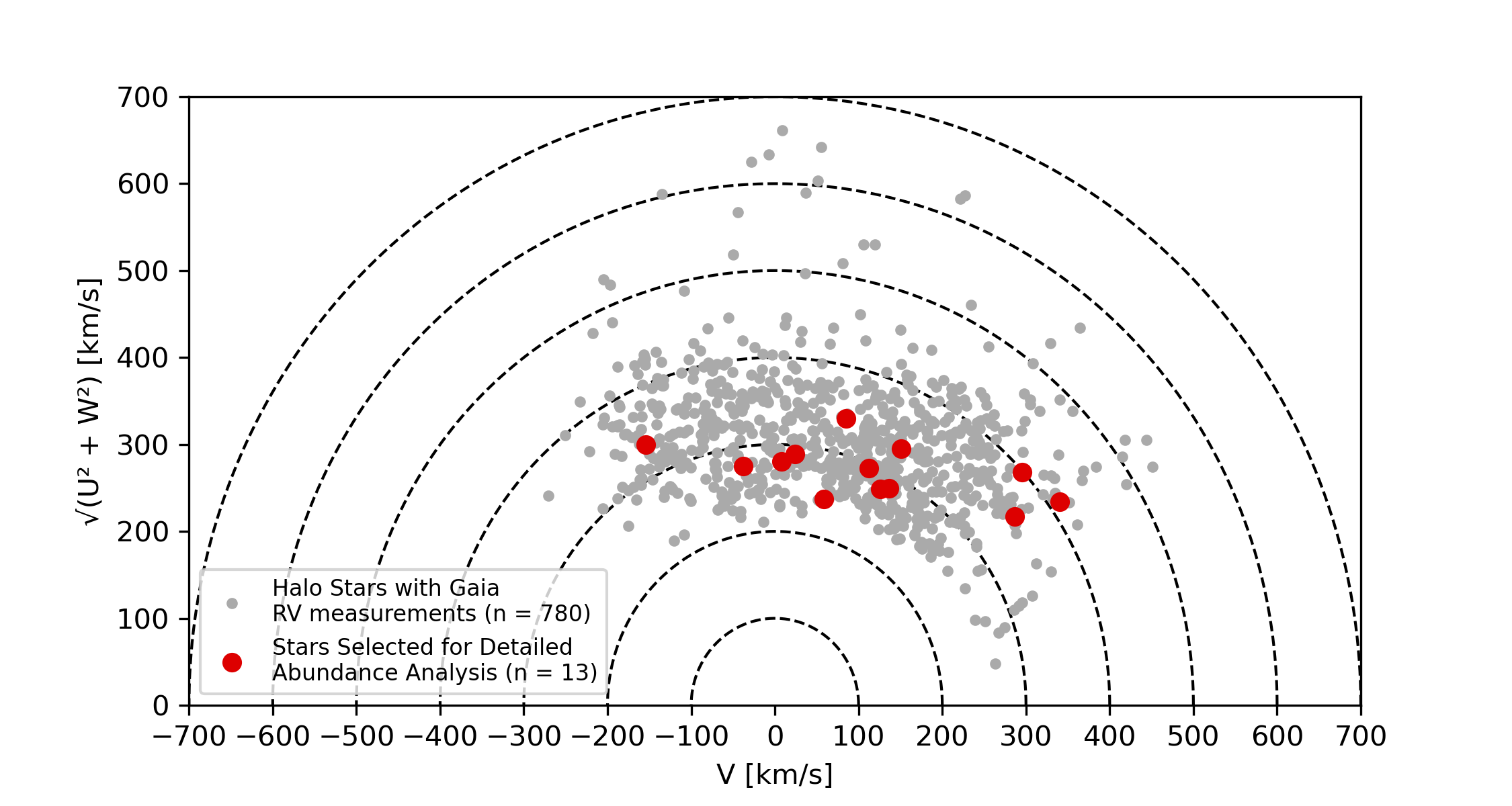}
 \caption{A projection of the stars' 3-dimensional space velocity onto a 2-dimensional plane, where the x-axis represents velocity in the direction of galactic rotation (V), and the y-axis represents the magnitude of the vector sum of radial velocity (U) and velocity perpendicular to the galactic plane (W). A star's total velocity is represented by its distance from the origin point (0,0). For this reason, iso-velocity lines are shown in gray for readability. All velocities are reported relative to the local standard of rest. Error bars are approximately 3 km/s along either axis and thus are not visible on the scale of the plot.}\label{fig:Toomre}
\end{figure*}

We calculated the Galactic space velocity~\citep{JohnsonSoderblom1987} for the 780 stars in the halo sample that had available Gaia DR2 radial velocity data and the 61 stars in the sample with APOGEE radial velocity data. Because radial velocity data are not used our selection procedure, we can further verify the validity of our selection with the galactic space velocity distribution of the subsample with radial velocity data. The Toomre diagram \citep{sandage1987} in Figure \ref{fig:Toomre} shows that the galactic space velocities of these stars are all above 200 km/s. We note that the distribution is consistent with that from the literature that studied metal-poor halo stars, which have $V_{\mathrm{tot}} > 180\ \mathrm{km}\,\mathrm{s}^{-1}$ \citep[e.g.,][]{nissen_schuster2010,schuster+2012}.

\subsection{Metallicity Distribution}
In addition to the 3-D kinematic verification of the sample, we can also directly measure the degree to which it is indeed metal-poor by appealing to independent metallicities from the literature, which have been compiled by the TIC.

\begin{figure}[ht!]\centering
 \includegraphics[width=\columnwidth]{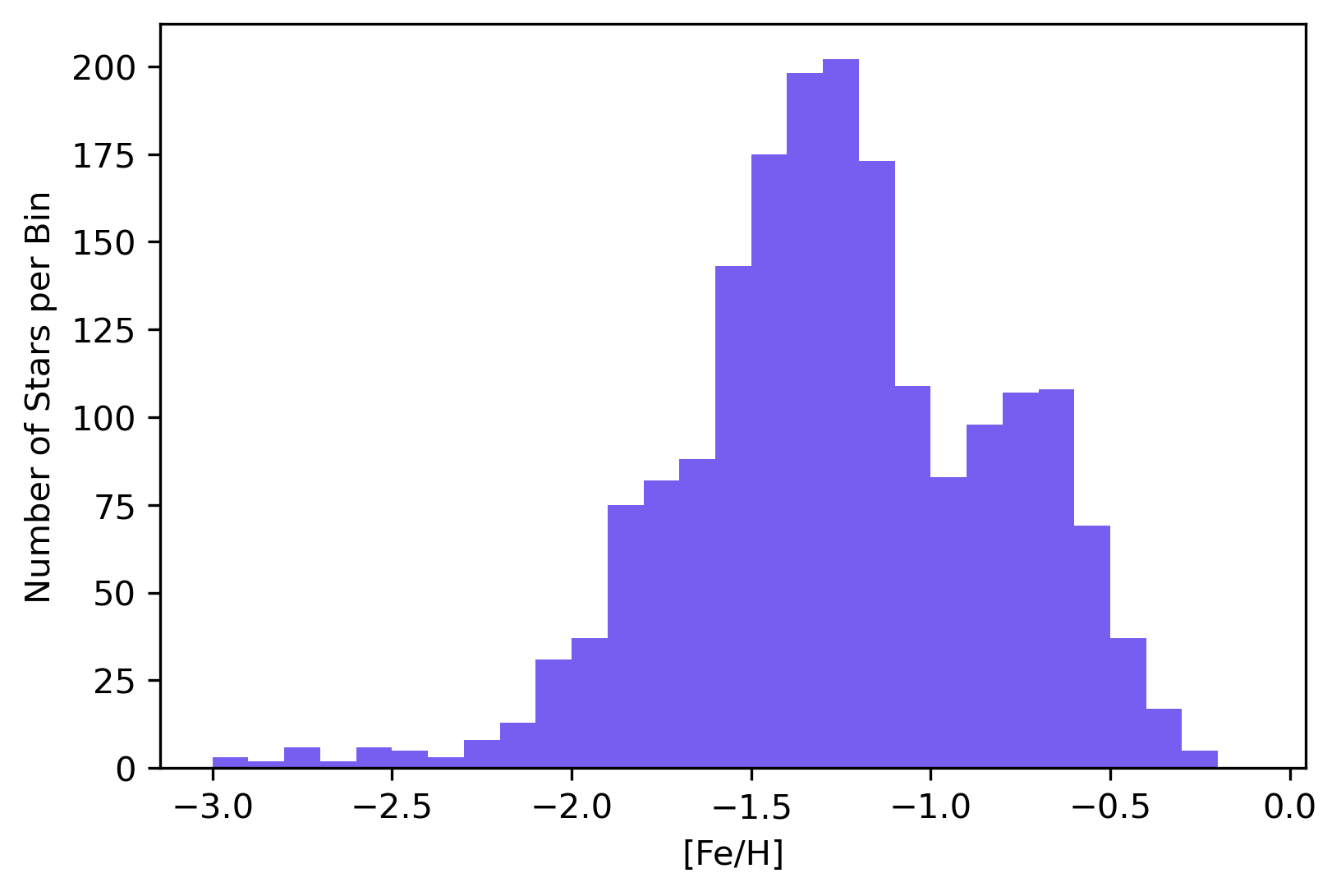}
 \caption{The distribution of TIC metallicity values (n = 1,897) of stars in the sample we selected. From the distribution, we estimate our complete sample to have a purity of roughly 70\%, as both distributions show 70\% of their stars having $\feh < -1.0$.}\label{fig:TIC}
\end{figure}

We analyzed the distribution of the TIC metallicities, which are available for roughly 11\% (n = 1,897) of the sample, yielded an overall purity estimate of $70\%$, the proportion of TIC metallicities which are below the low metallicity threshold of $\feh = 1.0$. We note that while the fraction of stars in the halo sample which have TIC metallicities is relatively small, the roughly 2,000 data points available should provide a metallicity distribution which is  sufficiently representative of that of the entire halo sample. The distribution of these TIC metallicities is shown in Figure~\ref{fig:TIC}.

It should be noted the metallicities in the TIC/CTL are compiled from literature spectroscopic values, and are therefore heterogeneous. In order of preference, metallicities for the catalogue are adopted from: SPOCS \citep{brewer+2016}; PASTEL \citep{soubiran+2016}; Gaia-ESO DR3 \citep{gilmore+2012}; TESS-HERMES DR1 \citep{sharma+2018}; GALAH DR2 \citep{buder+2018}; APOGEE-2 DR14 \citep{abolfathi+2018a}; LAMOST DR4 \citep{luo+2015}; RAVE DR5 \citep{kunder+2017}; and Geneva-Copenhagen DR3 \citep{holmberg_nordstrom_andersen2009}, as described in \citet{stassun+2019}. While the TIC metallicities may not be self-consistent and may have a few tenths dex uncertainties, it provides a representative distribution of the halo dwarf candidate sample metallicities.

\section{Observations of Halo Stars Showing Discrepancies in Stellar Properties}\label{Observations}

Some of our kinematically-selected halo stars have been previously known and studied~\citep{Reddy+2006,Sozzetti+2009,Boesgaard+2011}. However, there are discrepancies in derived stellar properties such as effective temperature and surface gravity, both among the literature, and when the literature is compared to the results of this paper. Table \ref{table:LitParams} shows effective temperature, surface gravity, and [Fe/H] values derived by previous studies. Notable anomalies include the following:~\begin{itemize}
\item BD+51 1696, the star measured by the greatest number of studies we compared with, shows significant spread in both \teff\ and log(g) between studies.
\item In many cases, \citet{Boesgaard+2011} derives surface gravities characteristic of post-main-sequence stars, despite the selection process of this paper being designed to limit the sample to main sequence dwarfs.
\end{itemize}

This motivates us to observe 13 previously-known halo stars using the PEPSI spectrograph~\citep{Strassmeier2015} at the Large Binocular Telescope (LBT, $2\times 8.4$\,m on Mt.\ Graham, Arizona, USA). The higher spectral resolution (R=120,000) and a homogeneous abundance analysis, in addition to the external constraints from the Gaia DR2 data, help to reconcile the discrepancies and lead to a more accurate determination of stellar properties. We detail the PEPSI observations and our abundance measurements below. 

LBT PEPSI observations were made on
March 25, May 15, June 23 and 25, 2019 UT with 200\,$\mu$m fiber ($R=130\,000$
and $1.75''$ on sky) in two spectral regions 4800 - 5441\,\AA\ and 6278 -
7419\,\AA\ (cross-dispersers (CD) 3 and 5) with 10--60\,min integration time
depending on the star brightness. A typical signal-to-noise ratio achieved is
350 in CD5 and 260 in CD3 for a star V=9.7 in one hour integration time with two
LBT mirrors. The whole sample has signal-to-noise ratio ranging from 100 to 600.

\begin{table*}
\centering
 \begin{tabular}{| c c c c |} 
 \hline
 Star & \teff\  (K)&  log(g) ([cm/s\textsuperscript{2}])& [Fe/H]   \\ [0.5ex] 
 \hline\hline
 BD+18 3423 &5943 (S), 5760 (R) & 4.43 (S), 4.59 (R) & -1.00 (S), -0.87 (R)\\ 
 \hline
 BD+20 2594 & 5886 (R) & 4.60 (R) & -0.94 (R)\\ 
 \hline
 BD+20 3603 & 5908 (B) & 3.61 (B) & -2.18 (B)\\ 
 \hline
 BD+25 1981 & 6745 (A) & 4.42 (A) & -1.45 (A)\\ 
 \hline
 BD+34 2476 & 6248 (B) & 3.72 (B) & -1.94 (B)\\ 
 \hline
 BD+36 2165 & 6052 (B) & 3.78 (B) & -1.71 (B)\\ 
 \hline
 BD+42 2667 & 5665 (R), 5793 (A) & 3.92 (R), 3.90 (A) & -1.34 (R), -1.48 (A)\\ 
 \hline
 BD+51 1696 & 5852 (B), 5315 (R), 5377 (A) & 4.19 (B), 4.74 (R), 3.90 (A) & -1.21 (B), -1.50 (R), -1.38 (A)\\ 
 \hline
 BD+75 839 & 5704 (R) & 4.13 (R) & -0.95 (R)\\ 
 \hline
 HD 64090 & 5384 (S), 5500 (B) & 4.70 (S), 4.73 (B) & -1.75 (S), -1.77 (B)\\ 
 \hline
 HD 108177 & 6105 (B) & 3.91 (B) & -1.77 (B)\\ 
 \hline
 HD 160693 & 5648 (R) & 4.48 (R) & -0.54 (R)\\ 
 \hline
 HD 194598 & 5875 (B) &  4.20 (B) & -1.23 (B)\\ 
 \hline
\end{tabular}
\caption{Effective temperatures derived by APOGEE DR16 \citep{APOGEEDR16}, \citet{Boesgaard+2011}, \citet{Reddy+2006}, and \citet{Sozzetti+2009}, as marked by the first letter of the citation.} \label{table:LitParams}
\end{table*}


\subsection{Data Reduction}

The data reduction is done using the Spectroscopic Data Systems (SDS) with its
pipeline adapted to the PEPSI data calibration flow and image specific
content. Its description is given in \citet{Strassmeier2018}.

The specific steps of image processing include bias subtraction and variance
estimation of the source images, super-master flat field correction for the CCD
spatial noise, \'echelle orders definition from the tracing flats, scattered
light subtraction, wavelength solution using ThAr images, optimal extraction
of image slices and cosmic ray removal, wavelength
calibration and merging slices in each order, normalization to the master flat
field spectrum to remove CCD fringes and blaze function, a global 2D fit to the
continuum of the normalized image, and rectification of all spectral orders in
the image to a 1D spectrum for a given cross-disperser.

The spectra from two sides of the telescope are averaged with weights into one
spectrum and corrected for the barycentric velocity of the Solar system. The
wavelength scale is preserved for each pixel as given by the wavelength solution
without rebinning.  The wavelength solution uses about 3000 ThAr lines and has
an uncertainty of the fit at the image center of 4\,m/s.

\section{Abundance Analysis}\label{Abundance Analysis}
\subsection{PEPSI spectra}
To determine the metallicities and stellar parameters of the stars observed with PEPSI, we first calculated the equivalent widths (EW) of absorption lines in each star's spectrum. This was done using an automated program which displays graphs and measurements from fits of a Voigt and Gaussian distribution, along with its direct measurement of the observed data. From this we chose the method which most closely fit the data trend. This manual screening allows for the best measurements to be kept for all lines, mitigating the effects of noise, contamination, and improper line fitting. If these effects are too great, the program allows us to discard the measurement from the final data.

To streamline this process, if all three EW measurements (Gaussian, Voigt, and raw) are within 20\% of each other, usually the result of a clean, isolated line feature, the program automatically takes the average of the three and uses the result as that line's EW. On the other hand, if both curve fitting functions fail, the line is automatically discarded without prompting the user. This latter scenario usually occurs where a line is weak enough that it is completely drowned out by noise and is indistinguishable from the continuum.

\begin{figure*}
\centering
\includegraphics[width=.8\paperwidth]{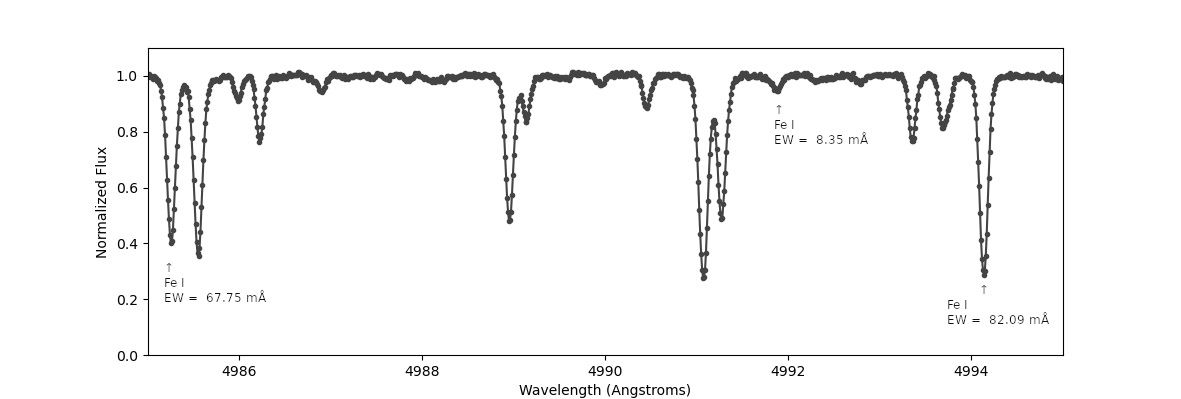}
  \caption{A range of the PEPSI B-band spectrum of HD 160693, with iron line features annotated with their corresponding equivalent width.}\label{fig:EQWs}
\end{figure*}

Line information was compiled from the NIST Atomic Spectra Database \citep{NIST_ASD}. A small section of the PEPSI spectrum for the star HD 160693 is shown in Figure \ref{fig:EQWs} with line features annotated with their corresponding species and equivalent width.

For each star, we chose to keep only lines for which EW $ > 5 \text{m\,\AA}$, to minimize the effects of noise on weak lines in our measurements. Furthermore, manual sigma-clipping based on each individual line's derived abundance was performed after the initial analysis to remove lingering outlying measurements, after which the process was repeated. This was done twice. A complete list of our final EW measurements can be found in Table \ref{table:EQWS}.

\subsection{Iterative MOOG Analysis}

To derive abundances and stellar atmospheric parameters, we used a program written by the authors which combines PyMOOGi\footnote{\url{https://github.com/madamow/pymoogi}} \citep{pyMOOGi}, which is a Python implementation of the Fortran code MOOG\footnote{\url{http://www.as.utexas.edu/\textasciitilde chris/moog.html}} \citep{Sneden1973}, with several Python functions designed to iteratively derive stellar parameters, abundances, and the uncertainties thereof.

We used the ATLAS9 model atmospheres computed for the APOGEE survey \citep{ATLAS9}, and the PyKMOD atmosphere interpolator\footnote{\url{https://github.com/kolecki4/PyKMOD}} to create the model stellar atmospheres for our analysis. To convert the resulting abundance values into quantities relative to solar, we took the solar iron abundance $\feh_{ \odot} = 7.48$ as stated in \citet{Palme+2014}.

\subsubsection{Temperature and Surface Gravity}\label{TandG}
We used Gaia DR2 \citep{Gaia2018}, 2MASS \citep{2MASS}, and WISE \citep{WISE} photometry compared against Dartmouth theoretical isochrones\footnote{\url{http://stellar.dartmouth.edu/models/}} \citep{Dotter+2007} to derive \teff\ and log(g). The photometry was queried from its respective databases using Astroquery, a module of the Astropy software library \citep{AstroPy}. We used distances computed by \citet{BailerJones2018} to convert these values from apparent to absolute magnitude.

The process is as follows: the star's photometric profile is compared against the synthetic photometry for each age and equivalent evolutionary phase (EEP) at a fixed [Fe/H] from the Dartmouth isochrones. From this comparison, a 2-dimensional grid of residuals is created in the age-EEP plane. Thus, each point on the grid represents the sum of the differences of magnitudes at each pass band between those observed of the star and those from the given point on the isochrone.

From the absolute minimum point on this grid, we define the area around this point for which the residuals are within 10\% of the minimum as valid points in the grid from which to extract \teff\ and log(g) values. The averages of the resulting temperature and gravity distributions are taken as the parameters derived from the given metallicity, with the standard deviation of the mean being taken as the uncertainty.

\subsubsection{Microturbulence}\label{mt}
To determine a value for the microturbulence parameter ($\xi$), we sought to remove the correlation between reduced equivalent width (REW) and abundance for \ion{Fe}{1} lines. This was done by calculating the slope of this correlation for a fixed set of microturbulence values (ranging from 0 to 4 km/s), and linearly interpolating the microturbulence value at which the resulting slope is 0.

In the case where this process did not succeed (for example, if there is no point where the slope value crosses 0 along the range of microturbulence values), we chose to follow the convention used by \citet{Boesgaard+2011}, which is to simply set microturbulence equal to 1.5 km/s.

\subsubsection{Details of the Iteration Program}\label{deets}

The process first assumes a metallicity of $\feh = -1.0$. From here, it gets \teff\ and log(g) values according to Section \ref{TandG}, and interpolates a model atmosphere from the grid with these photometric parameters and a microturbulence set to 1.5 km/s. Then, it runs MOOG with this model atmosphere and the line list for a given star.

It then reads the MOOG output, gathering the metallicity data and a value for [Fe/H], and then repeats, now using isochrones of the output [Fe/H] to derive new parameters for the model atmosphere. Convergence is reached and the process ends when the metallicity output by MOOG is the same (to within 0.1 dex, the precision of the isochrone grid) as that used to derive that run's stellar parameters.

Note that the initial metallicity choice is arbitrary. It is simply used as an initial guess for the iteration process, which lasts a variable amount of time based on how accurate this initial guess is.

After the process converges on a solution for [Fe/H], we calculate the microturbulence as outlined in Section \ref{mt}, re-deriving stellar parameters as necessary as the microturbulence affects the metallicity. At this point, the program proceeds with the uncertainty analysis.

\subsection{Uncertainty Analysis}

Uncertainty in the microturbulence parameter was determined by perturbing the microturbulence until the slope of the REW-abundance correlation fell outside of the uncertainty range.

All other uncertainties, including those for \teff, log(g), and abundances were calculated using iteration of the process outlined in Section 3.2 of \citet{Epstein+2010}. The method used by those authors creates a matrix of partial derivatives of parameters with respect to one another and uses various equations to account for the effects of uncertainties in each parameter on the uncertainties of all the others.

Since, by the methodology outlined in Section \ref{TandG}, the uncertainty in metallicity can affect the uncertainty in \teff\ and log(g), and by the methodology outlined here, the reverse is also true, we are required to iterate these calculations repeatedly, using the output uncertainties from one run as the input uncertainties of the next.

Initial uncertainties were simply taken as the standard deviation of the mean value derived for each parameter, and the calculations taken from Epstein et al. were repeated until the [Fe/H] uncertainty was changed by less than 0.01 dex from one iteration to the next.

\subsection{Results}
\begin{table*}[ht]
\tablenum{4}
\centering
 \begin{tabular}{|c c c c c c|} 
 \hline
 Star & \teff\ (K) & log(g) ([cm/s\textsuperscript{2}]) & $\xi$ (km/s) & [\ion{Fe}{1}/H] & [\ion{Fe}{2}/H]\\  
 \hline\hline
 BD+18 3423 & 6204 $\pm$ 70 & 4.19 $\pm$ 0.01 & 0.86 $\pm$ 0.21 & -0.874 $\pm$ 0.099 & -0.893 $\pm$ 0.103\\ 
 \hline
 BD+20 2594 & 6160 $\pm$ 40 & 4.29 $\pm$ 0.01 & 0.42 $\pm$ 0.29 & -0.838 $\pm$ 0.067 & -0.886 $\pm$ 0.079\\ 
 \hline
 BD+20 3603 & 6544 $\pm$ 20 & 4.32 $\pm$ 0.01 & 1.52 $\pm$ 0.28 & -1.982 $\pm$ 0.116 & -2.088 $\pm$ 0.140\\ 
 \hline
 BD+25 1981 & 6774 $\pm$ 20 & 4.18 $\pm$ 0.02 & 1.52 $\pm$ 0.24 & -1.551 $\pm$ 0.140 & -1.387 $\pm$ 0.150\\ 
 \hline
 BD+34 2476 & 6595 $\pm$ 40 & 4.11 $\pm$ 0.01 & 1.34 $\pm$ 0.21 & -1.884 $\pm$ 0.110 & -1.941 $\pm$ 0.145\\ 
 \hline
 BD+36 2165 & 6470 $\pm$ 20 & 4.19 $\pm$ 0.01 & 0.86 $\pm$ 0.23 & -1.359 $\pm$ 0.053 & -1.411 $\pm$ 0.046\\ 
 \hline
 BD+42 2667 & 6314 $\pm$ 30 & 4.32 $\pm$ 0.01 & 0.56 $\pm$ 0.27 & -1.265 $\pm$ 0.086 & -1.328 $\pm$ 0.091\\ 
 \hline
 BD+51 1696 & 5793 $\pm$ 20 & 4.57 $\pm$ 0.01 & 1.5 $\pm$ 0.71 & -1.424 $\pm$ 0.084 & -1.416 $\pm$ 0.096\\ 
 \hline
 BD+75 839 & 6320 $\pm$ 70 & 4.21 $\pm$ 0.05 & 1.10 $\pm$ 0.13 & -0.769 $\pm$ 0.158 & -1.151 $\pm$ 0.164\\ 
 \hline
 HD 64090 & 5607 $\pm$ 10 & 4.67 $\pm$ 0.01 & 1.5 $\pm$ 0.78 & -1.650 $\pm$ 0.074 & -1.846 $\pm$ 0.107\\ 
 \hline
 HD 108177 & 6410 $\pm$ 40 & 4.33 $\pm$ 0.01 & 0.59 $\pm$ 0.30 & -1.479 $\pm$ 0.113 & -1.502 $\pm$ 0.160\\ 
 \hline
 HD 160693 & 5951 $\pm$ 30 & 4.29 $\pm$ 0.01 & 1.5 $\pm$ 0.52 & -0.610 $\pm$ 0.052 & -0.613 $\pm$ 0.078\\ 
 \hline
 HD 194598 & 6243 $\pm$ 80 & 4.36 $\pm$ 0.02 & 1.5 $\pm$ 0.40 & -1.131 $\pm$ 0.192 & -1.243 $\pm$ 0.209\\ 
 \hline
\end{tabular}
\caption{Stellar Parameters of Selected Sample Stars} \label{table:Stellar Parameters}
\end{table*}

We found the stars to fall within a temperature range of $ 5600 \leq $ \teff\ $ \leq 6800$ and a surface gravity range of $ 4.1 \leq log(g) \leq 4.7 $.  Detailed stellar parameter and abundance results for each star can be found in Table \ref{table:Stellar Parameters}. The stars also fall well within the expected metallicity values given their halo classification, as we found them to fall within the range $-2.0 < \feh < -0.6$.

\begin{figure}[h]
\centering
 \includegraphics[width=1.1\linewidth]{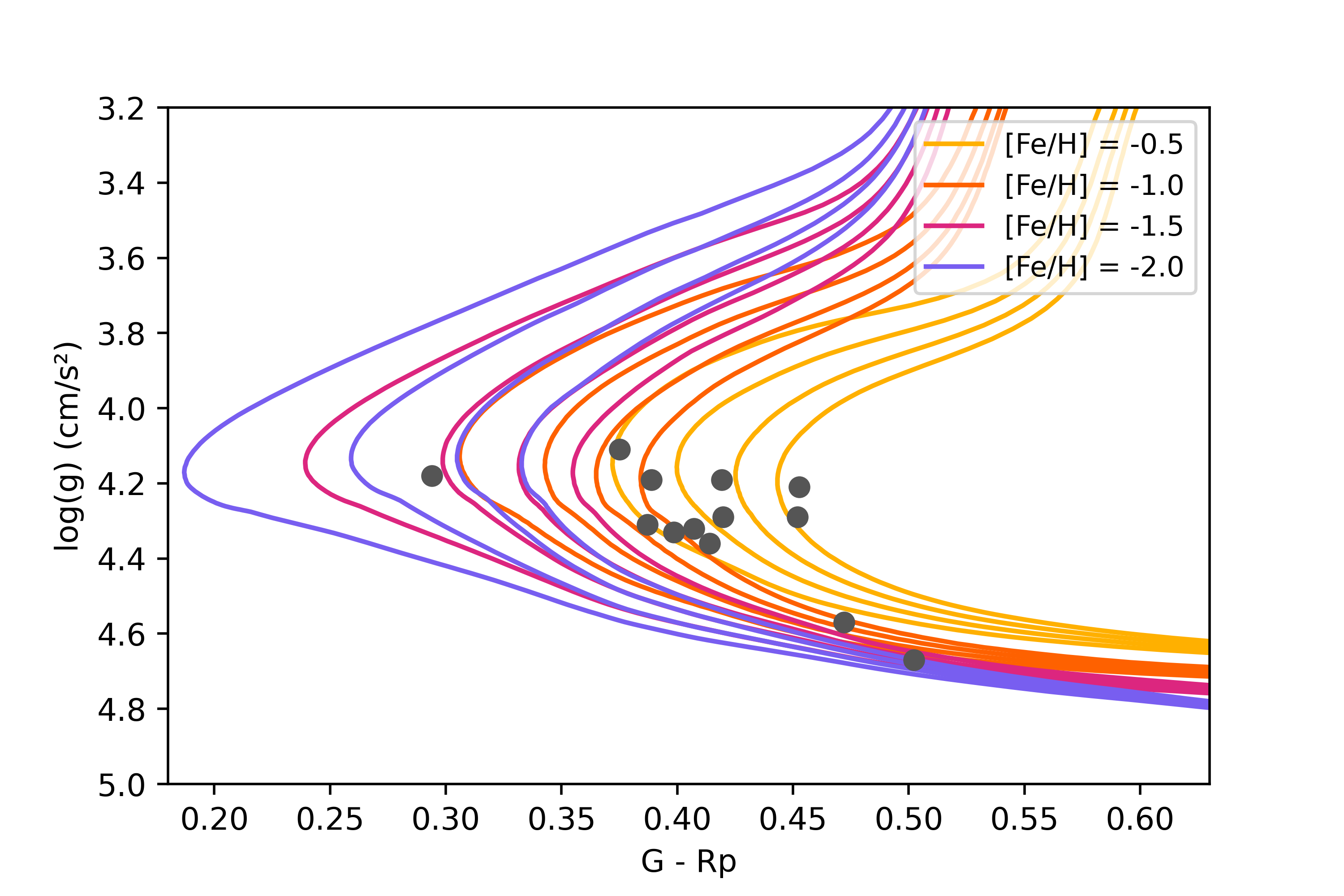}
 \caption{A plot of log(g) versus $(G-Rp)$ color, overlayed with isochrones of 6, 8, 10, and 12 Gyr for each metallicity.}\label{fig:HRcomp}
\end{figure}

\section{Discussion}\label{Discussion}



\subsection{Using Photometric vs Spectroscopic Parameters}\label{AnalysisMethods}

One method of deriving stellar parameters involves using the spectroscopic abundance analysis, adjusting them until the following conditions are met:

1. Effective Temperature: The star should be in excitation equilibrium, i.e. the correlation between abundance derived from each line and excitation potential (EP) should be removed.

2. Surface Gravity: The star should be in ionization balance, i.e. the abundances of \ion{Fe}{1} and \ion{Fe}{2} should be within 1$\sigma$ of each other.

3. Microturbulence: There should be no correlation between the abundance derived from each line and reduced equivalent width ($log(\frac{\text{EW}}{\lambda})$).

We followed this method for microturbulence, but our \teff\ and log(g) values were derived photometrically from isochrones. However, for every star except BD+75 839, the resulting log(g) led to ionization balance.

Also, the resulting \teff\ led to excitation equilibrium for many of the stars, such that the slope of the EP-abundance regression line was equal to 0 at the 1-sigma level for four stars, and at the 3-sigma for seven, where sigma is defined as the standard deviation of the linear regression fit assuming normal distribution of residuals.

Two stars (HD 64090 and HD 160693) featured slopes which were 4-sigma away from zero. Notably however, these two stars were also unable to achieve microturbulence convergence.

\subsubsection{Comparison with Previous Results}

For the six stars in common with~\citet{Reddy+2006}, the mutual discrepancy between metallicities is on average 0.06 dex, which we consider to be reasonably identical within a margin of error. Their process of determining stellar parameters was similar to ours in that it also does not rely on spectroscopy for \teff\ and log(g). They use $(b-y)$ and $(V-K_\mathrm{s})$ to derive \teff, whereas we use magnitudes, rather than colors, and use $G$, $Bp$, $Rp$, $J$, $H$, $K_\mathrm{s}$, $W1$, $W2$, $W3$, and $W4$.

For surface gravities, they make use of Hipparcos astrometry whereas we use the same photometric method for log(g) as was used for \teff. And lastly, for microturbulence, they use a previously-calculated relation between other parameters and $\xi$ for metal-poor dwarfs, where we in this case use the spectroscopic approach for microturbulence.

We also compared our results with those for the seven stars our sample shares with \citet{Boesgaard+2011}. In the paper, they determined the stellar parameters spectroscopically according to the conditions in Section \ref{AnalysisMethods}. Their metallicities are systematically lower than ours by an average of 0.19 dex, except in the case of BD+51 1696, where theirs is higher by 0.23 dex.

Although similar stellar parameter convergence conditions (ionization balance, excitation equilibrium, microturbulence convergence) were met to a degree of uncertainty in both papers, it can be seen in Tables \ref{table:LitParams} and \ref{table:Stellar Parameters} that the resulting parameters are in many cases quite different, especially surface gravity.

From these discrepancies we can assume the importance of the additional photometric constraints on the parameters.

Based on the available information, it seems that the spectroscopic method can lead to cases where there are multiple points in the \teff-log(g)-$\xi$ parameter space where the convergence conditions are met. This leaves the chance that, if additional constraints are not used, any given analysis may not necessarily converge on the correct point, leading to parameters which conflict with other stellar data.

\subsection{The Negligibility of non-LTE Corrections}
We investigated the magnitude of the effects of the LTE assumption on our final abundance measurements. We accessed data from \citet{Bergemann+2012} via a web tool by \citet{NLTE_MPIA}, and discovered that all corrections for lines we tested were within our metallicity uncertainty thresholds, thus making them negligible corrections to our measurements.

\section{Summary and Conclusion}

In this paper we discuss our selection process for metal-poor halo stars, detailing kinematic and photometric criteria resulting in a final sample of $\sim16,940$ stars, where $\feh < -1.0$ for roughly $70\%$ of the stars included based on comparison to literature metallicities.

We also present a re-analysis of 13 halo stars from the literature using new observations taken with the high-resolution PEPSI spectrograph, attempting to rectify discrepancies of stellar properties in the literature. We measure the metallicity of these stars to an accuracy of roughly $\sigma(\feh) = 0.1~\text{dex}$. In this process, we also used Gaia, 2MASS, and WISE photometry to derive accurate effective temperature and surface gravity values using Dartmouth theoretical isochrones.

In summary, given the overall fidelity of our sample, the halo dwarf candidates presented here will prove to be useful targets for planet studies in metal-poor host systems with TESS. An analysis of the sample in the context of planet occurrence rates in this metal-poor regime is described in Boley et al. 2021, the companion to this paper.

\acknowledgments
JCZ supported by an NSF Astronomy and Astrophysics Postdoctoral Fellowship under award AST-2001869.

This work has made use of data from the European Space Agency (ESA) mission
{\it Gaia} (\url{https://www.cosmos.esa.int/gaia}), processed by the {\it Gaia}
Data Processing and Analysis Consortium (DPAC,
\url{https://www.cosmos.esa.int/web/gaia/dpac/consortium}). Funding for the DPAC
has been provided by national institutions, in particular the institutions
participating in the {\it Gaia} Multilateral Agreement. 

This research made use of Astropy,\footnote{http://www.astropy.org} a community-developed core Python package for Astronomy \citep{astropy:2013, astropy:2018}. 

PEPSI was made possible by funding through the State of Brandenburg (MWFK) and the German Federal Ministry of Education and Research (BMBF) through their Verbundforschung grants 05AL2BA1/3 and 05A08BAC.

The LBT is an international collaboration among institutions in the
United States, Italy and Germany. LBT Corporation partners are: The
University of Arizona on behalf of the Arizona Board of Regents;
Istituto Nazionale di Astrofisica, Italy; LBT Beteiligungsgesellschaft,
Germany, representing the Max-Planck Society, The Leibniz Institute for
Astrophysics Potsdam, and Heidelberg University; The Ohio State
University, representing OSU, University of Notre Dame, University of
Minnesota and University of Virginia.

Funding for the Sloan Digital Sky Survey IV has been provided by the  Alfred P. Sloan Foundation, the U.S. Department of Energy Office of Science, and the Participating Institutions. 

SDSS-IV acknowledges support and resources from the Center for High Performance Computing  at the University of Utah. The SDSS website is www.sdss.org.

SDSS-IV is managed by the Astrophysical Research Consortium for the Participating Institutions of the SDSS Collaboration including the Brazilian Participation Group, the Carnegie Institution for Science, Carnegie Mellon University, Center for Astrophysics | Harvard \& Smithsonian, the Chilean Participation Group, the French Participation Group, Instituto de Astrof\'isica de Canarias, The Johns Hopkins University, Kavli Institute for the Physics and Mathematics of the Universe (IPMU) / University of Tokyo, the Korean Participation Group, Lawrence Berkeley National Laboratory, Leibniz Institut f\"ur Astrophysik Potsdam (AIP),  Max-Planck-Institut f\"ur Astronomie (MPIA Heidelberg), Max-Planck-Institut f\"ur Astrophysik (MPA Garching), Max-Planck-Institut f\"ur Extraterrestrische Physik (MPE), National Astronomical Observatories of China, New Mexico State University, New York University, University of Notre Dame, Observat\'ario Nacional / MCTI, The Ohio State University, Pennsylvania State University, Shanghai Astronomical Observatory, United Kingdom Participation Group, Universidad Nacional Aut\'onoma de M\'exico, University of Arizona, University of Colorado Boulder, University of Oxford, University of Portsmouth, University of Utah, University of Virginia, University of Washington, University of Wisconsin, Vanderbilt University, and Yale University.

\software{APOGEE data reduction pipeline \citep{Nidever}, ASPCAP \citep{GarciaPerez}), Astropy \citep{AstroPy}, MOOG \citep{Sneden1973}, PyMOOGi \citep{pyMOOGi}, SciPy \citep{Scipy}, SDS \citep{Strassmeier2018}}

\bibliography{bib}

\begin{thebibliography}{}
\expandafter\ifx\csname natexlab\endcsname\relax\def\natexlab#1{#1}\fi
\providecommand{\url}[1]{\href{#1}{#1}}
\providecommand{\dodoi}[1]{doi:~\href{http://doi.org/#1}{\nolinkurl{#1}}}
\providecommand{\doeprint}[1]{\href{http://ascl.net/#1}{\nolinkurl{http://ascl.net/#1}}}
\providecommand{\doarXiv}[1]{\href{https://arxiv.org/abs/#1}{\nolinkurl{https://arxiv.org/abs/#1}}}

\bibitem[{{Abolfathi} {et~al.}(2018){Abolfathi}, {Aguado}, {Aguilar}, {Allende
  Prieto}, {Almeida}, {Ananna}, {Anders}, {Anderson}, {Andrews}, {Anguiano}, \&
  et~al.}]{abolfathi+2018a}
{Abolfathi}, B., {Aguado}, D.~S., {Aguilar}, G., {et~al.} 2018, \apjs, 235, 42

\bibitem[{{Adamow}(2017)}]{pyMOOGi}
{Adamow}, M.~M. 2017, in American Astronomical Society Meeting Abstracts, Vol.
  230, American Astronomical Society Meeting Abstracts \#230, 216.07

\bibitem[{{Adibekyan} {et~al.}(2012){Adibekyan}, {Santos}, {Sousa},
  {Israelian}, {Delgado Mena}, {Gonz{\'a}lez Hern{\'a}ndez}, {Mayor}, {Lovis},
  \& {Udry}}]{Adibekyan2012}
{Adibekyan}, V.~Z., {Santos}, N.~C., {Sousa}, S.~G., {et~al.} 2012, A\&A, 543,
  A89, \dodoi{10.1051/0004-6361/201219564}

\bibitem[{{Anglada-Escud{\'e}} {et~al.}(2014){Anglada-Escud{\'e}}, {Arriagada},
  {Tuomi}, {Zechmeister}, {Jenkins}, {Ofir}, {Dreizler}, {Gerlach}, {Marvin},
  {Reiners}, {Jeffers}, {Butler}, {Vogt}, {Amado},
  {Rodr{\'{\i}}guez-L{\'o}pez}, {Berdi{\~n}as}, {Morin}, {Crane}, {Shectman},
  {Thompson}, {D{\'{\i}}az}, {Rivera}, {Sarmiento}, \& {Jones}}]{Escude2014}
{Anglada-Escud{\'e}}, G., {Arriagada}, P., {Tuomi}, M., {et~al.} 2014, MNRAS,
  443, L89, \dodoi{10.1093/mnrasl/slu076}

\bibitem[{{Astropy Collaboration} {et~al.}(2013){Astropy Collaboration},
  {Robitaille}, {Tollerud}, {Greenfield}, {Droettboom}, {Bray}, {Aldcroft},
  {Davis}, {Ginsburg}, {Price-Whelan}, {Kerzendorf}, {Conley}, {Crighton},
  {Barbary}, {Muna}, {Ferguson}, {Grollier}, {Parikh}, {Nair}, {Unther},
  {Deil}, {Woillez}, {Conseil}, {Kramer}, {Turner}, {Singer}, {Fox}, {Weaver},
  {Zabalza}, {Edwards}, {Azalee Bostroem}, {Burke}, {Casey}, {Crawford},
  {Dencheva}, {Ely}, {Jenness}, {Labrie}, {Lim}, {Pierfederici}, {Pontzen},
  {Ptak}, {Refsdal}, {Servillat}, \& {Streicher}}]{astropy:2013}
{Astropy Collaboration}, {Robitaille}, T.~P., {Tollerud}, E.~J., {et~al.} 2013,
  \aap, 558, A33, \dodoi{10.1051/0004-6361/201322068}

\bibitem[{{Astropy Collaboration} {et~al.}(2018){Astropy Collaboration},
  {Price-Whelan}, {Sip{H{o}}cz}, {G{"u}nther}, {Lim}, {Crawford}, {Conseil},
  {Shupe}, {Craig}, {Dencheva}, {Ginsburg}, {Vand erPlas}, {Bradley},
  {P{'e}rez-Su{'a}rez}, {de Val-Borro}, {Aldcroft}, {Cruz}, {Robitaille},
  {Tollerud}, {Ardelean}, {Babej}, {Bach}, {Bachetti}, {Bakanov}, {Bamford},
  {Barentsen}, {Barmby}, {Baumbach}, {Berry}, {Biscani}, {Boquien}, {Bostroem},
  {Bouma}, {Brammer}, {Bray}, {Breytenbach}, {Buddelmeijer}, {Burke},
  {Calderone}, {Cano Rodr{'i}guez}, {Cara}, {Cardoso}, {Cheedella}, {Copin},
  {Corrales}, {Crichton}, {D'Avella}, {Deil}, {Depagne}, {Dietrich}, {Donath},
  {Droettboom}, {Earl}, {Erben}, {Fabbro}, {Ferreira}, {Finethy}, {Fox},
  {Garrison}, {Gibbons}, {Goldstein}, {Gommers}, {Greco}, {Greenfield},
  {Groener}, {Grollier}, {Hagen}, {Hirst}, {Homeier}, {Horton}, {Hosseinzadeh},
  {Hu}, {Hunkeler}, {Ivezi{'c}}, {Jain}, {Jenness}, {Kanarek}, {Kendrew},
  {Kern}, {Kerzendorf}, {Khvalko}, {King}, {Kirkby}, {Kulkarni}, {Kumar},
  {Lee}, {Lenz}, {Littlefair}, {Ma}, {Macleod}, {Mastropietro}, {McCully},
  {Montagnac}, {Morris}, {Mueller}, {Mumford}, {Muna}, {Murphy}, {Nelson},
  {Nguyen}, {Ninan}, {N{"o}the}, {Ogaz}, {Oh}, {Parejko}, {Parley}, {Pascual},
  {Patil}, {Patil}, {Plunkett}, {Prochaska}, {Rastogi}, {Reddy Janga},
  {Sabater}, {Sakurikar}, {Seifert}, {Sherbert}, {Sherwood-Taylor}, {Shih},
  {Sick}, {Silbiger}, {Singanamalla}, {Singer}, {Sladen}, {Sooley},
  {Sornarajah}, {Streicher}, {Teuben}, {Thomas}, {Tremblay}, {Turner},
  {Terr{'o}n}, {van Kerkwijk}, {de la Vega}, {Watkins}, {Weaver}, {Whitmore},
  {Woillez}, {Zabalza}, \& {Astropy Contributors}}]{astropy:2018}
{Astropy Collaboration}, {Price-Whelan}, A.~M., {Sip{H{o}}cz}, B.~M., {et~al.}
  2018, aj, 156, 123, \dodoi{10.3847/1538-3881/aabc4f}

\bibitem[{{Bailer-Jones} {et~al.}(2018){Bailer-Jones}, {Rybizki}, {Fouesneau},
  {Mantelet}, \& {Andrae}}]{BailerJones2018}
{Bailer-Jones}, C.~A.~L., {Rybizki}, J., {Fouesneau}, M., {Mantelet}, G., \&
  {Andrae}, R. 2018, \aj, 156, 58, \dodoi{10.3847/1538-3881/aacb21}

\bibitem[{{Bashi} \& {Zucker}(2019)}]{Bashi2019}
{Bashi}, D., \& {Zucker}, S. 2019, \aj, 158, 61,
  \dodoi{10.3847/1538-3881/ab27c9}

\bibitem[{{Bergemann} {et~al.}(2012){Bergemann}, {Lind}, {Collet}, {Magic}, \&
  {Asplund}}]{Bergemann+2012}
{Bergemann}, M., {Lind}, K., {Collet}, R., {Magic}, Z., \& {Asplund}, M. 2012,
  \mnras, 427, 27, \dodoi{10.1111/j.1365-2966.2012.21687.x}

\bibitem[{{Boesgaard} {et~al.}(2011){Boesgaard}, {Rich}, {Levesque}, \&
  {Bowler}}]{Boesgaard+2011}
{Boesgaard}, A.~M., {Rich}, J.~A., {Levesque}, E.~M., \& {Bowler}, B.~P. 2011,
  \apj, 743, 140, \dodoi{10.1088/0004-637X/743/2/140}

\bibitem[{{Bond} {et~al.}(2010){Bond}, {Ivezi{\'c}}, {Sesar}, {Juri{\'c}},
  {Munn}, {Kowalski}, {Loebman}, {Ro{\v{s}}kar}, {Beers}, {Dalcanton},
  {Rockosi}, {Yanny}, {Newberg}, {Allende Prieto}, {Wilhelm}, {Lee},
  {Sivarani}, {Majewski}, {Norris}, {Bailer-Jones}, {Re Fiorentin}, {Schlegel},
  {Uomoto}, {Lupton}, {Knapp}, {Gunn}, {Covey}, {Allyn Smith}, {Miknaitis},
  {Doi}, {Tanaka}, {Fukugita}, {Kent}, {Finkbeiner}, {Quinn}, {Hawley},
  {Anderson}, {Kiuchi}, {Chen}, {Bushong}, {Sohi}, {Haggard}, {Kimball},
  {McGurk}, {Barentine}, {Brewington}, {Harvanek}, {Kleinman}, {Krzesinski},
  {Long}, {Nitta}, {Snedden}, {Lee}, {Pier}, {Harris}, {Brinkmann}, \&
  {Schneider}}]{bond+2010}
{Bond}, N.~A., {Ivezi{\'c}}, {\v{Z}}., {Sesar}, B., {et~al.} 2010, \apj, 716, 1

\bibitem[{{Brewer} {et~al.}(2016){Brewer}, {Fischer}, {Valenti}, \&
  {Piskunov}}]{brewer+2016}
{Brewer}, J.~M., {Fischer}, D.~A., {Valenti}, J.~A., \& {Piskunov}, N. 2016,
  \apjs, 225, 32

\bibitem[{{Buder} {et~al.}(2018){Buder}, {Asplund}, {Duong}, {Kos}, {Lind},
  {Ness}, {Sharma}, {Bland-Hawthorn}, {Casey}, {De Silva}, {D'Orazi},
  {Freeman}, {Lewis}, {Lin}, {Martell}, {Schlesinger}, {Simpson}, {Zucker},
  {Zwitter}, {Amarsi}, {Anguiano}, {Carollo}, {Casagrande}, {{\v C}otar},
  {Cottrell}, {Da Costa}, {Gao}, {Hayden}, {Horner}, {Ireland}, {Kafle},
  {Munari}, {Nataf}, {Nordlander}, {Stello}, {Ting}, {Traven}, {Watson},
  {Wittenmyer}, {Wyse}, {Yong}, {Zinn}, \& {{\v Z}erjal}}]{buder+2018}
{Buder}, S., {Asplund}, M., {Duong}, L., {et~al.} 2018, \mnras, 478, 4513

\bibitem[{{Campante} {et~al.}(2015){Campante}, {Barclay}, {Swift}, {Huber},
  {Adibekyan}, {Cochran}, {Burke}, {Isaacson}, {Quintana}, {Davies}, {Silva
  Aguirre}, {Ragozzine}, {Riddle}, {Baranec}, {Basu}, {Chaplin},
  {Christensen-Dalsgaard}, {Metcalfe}, {Bedding}, {Handberg}, {Stello},
  {Brewer}, {Hekker}, {Karoff}, {Kolbl}, {Law}, {Lundkvist}, {Miglio}, {Rowe},
  {Santos}, {Van Laerhoven}, {Arentoft}, {Elsworth}, {Fischer}, {Kawaler},
  {Kjeldsen}, {Lund}, {Marcy}, {Sousa}, {Sozzetti}, \& {White}}]{Campante2015}
{Campante}, T.~L., {Barclay}, T., {Swift}, J.~J., {et~al.} 2015, ApJ, 799, 170,
  \dodoi{10.1088/0004-637X/799/2/170}

\bibitem[{{Carney} \& {Latham}(1987)}]{Carney+Latham}
{Carney}, B.~W., \& {Latham}, D.~W. 1987, \aj, 93, 116, \dodoi{10.1086/114292}

\bibitem[{{Choi} {et~al.}(2016){Choi}, {Dotter}, {Conroy}, {Cantiello},
  {Paxton}, \& {Johnson}}]{choi+2016a}
{Choi}, J., {Dotter}, A., {Conroy}, C., {et~al.} 2016, \apj, 823, 102

\bibitem[{{Dotter}(2016)}]{dotter+2016a}
{Dotter}, A. 2016, \apjs, 222, 8

\bibitem[{{Dotter} {et~al.}(2007){Dotter}, {Chaboyer}, {Jevremovi{\'c}},
  {Baron}, {Ferguson}, {Sarajedini}, \& {Anderson}}]{Dotter+2007}
{Dotter}, A., {Chaboyer}, B., {Jevremovi{\'c}}, D., {et~al.} 2007, \aj, 134,
  376, \dodoi{10.1086/517915}

\bibitem[{{Eggen} {et~al.}(1962){Eggen}, {Lynden-Bell}, \& {Sandage}}]{els}
{Eggen}, O.~J., {Lynden-Bell}, D., \& {Sandage}, A.~R. 1962, \apj, 136, 748,
  \dodoi{10.1086/147433}

\bibitem[{{Epstein} {et~al.}(2010){Epstein}, {Johnson}, {Dong}, {Udalski},
  {Gould}, \& {Becker}}]{Epstein+2010}
{Epstein}, C.~R., {Johnson}, J.~A., {Dong}, S., {et~al.} 2010, \apj, 709, 447,
  \dodoi{10.1088/0004-637X/709/1/447}

\bibitem[{{Faria} {et~al.}(2016){Faria}, {Santos}, {Figueira}, {Mortier},
  {Dumusque}, {Boisse}, {Lo Curto}, {Lovis}, {Mayor}, {Melo}, {Pepe}, {Queloz},
  {Santerne}, {S{\'e}gransan}, {Sousa}, {Sozzetti}, \& {Udry}}]{Faria2016}
{Faria}, J.~P., {Santos}, N.~C., {Figueira}, P., {et~al.} 2016, A\&A, 589, A25,
  \dodoi{10.1051/0004-6361/201527522}

\bibitem[{{Fischer} \& {Valenti}(2005)}]{Fischer2005}
{Fischer}, D.~A., \& {Valenti}, J. 2005, ApJ, 622, 1102, \dodoi{10.1086/428383}

\bibitem[{{Gaia Collaboration} {et~al.}(2016){Gaia Collaboration}, {Prusti},
  {de Bruijne}, {Brown}, {Vallenari}, {Babusiaux}, {Bailer-Jones}, {Bastian},
  {Biermann}, {Evans}, {Eyer}, {Jansen}, {Jordi}, {Klioner}, {Lammers},
  {Lindegren}, {Luri}, {Mignard}, {Milligan}, {Panem}, {Poinsignon},
  {Pourbaix}, {Randich}, {Sarri}, {Sartoretti}, {Siddiqui}, {Soubiran},
  {Valette}, {van Leeuwen}, {Walton}, {Aerts}, {Arenou}, {Cropper}, {Drimmel},
  {H{\o}g}, {Katz}, {Lattanzi}, {O'Mullane}, {Grebel}, {Holland}, {Huc},
  {Passot}, {Bramante}, {Cacciari}, {Casta{\~n}eda}, {Chaoul}, {Cheek}, {De
  Angeli}, {Fabricius}, {Guerra}, {Hern{\'a}ndez}, {Jean-Antoine-Piccolo},
  {Masana}, {Messineo}, {Mowlavi}, {Nienartowicz}, {Ord{\'o}{\~n}ez-Blanco},
  {Panuzzo}, {Portell}, {Richards}, {Riello}, {Seabroke}, {Tanga},
  {Th{\'e}venin}, {Torra}, {Els}, {Gracia-Abril}, {Comoretto},
  {Garcia-Reinaldos}, {Lock}, {Mercier}, {Altmann}, {Andrae}, {Astraatmadja},
  {Bellas-Velidis}, {Benson}, {Berthier}, {Blomme}, {Busso}, {Carry},
  {Cellino}, {Clementini}, {Cowell}, {Creevey}, {Cuypers}, {Davidson}, {De
  Ridder}, {de Torres}, {Delchambre}, {Dell'Oro}, {Ducourant}, {Fr{\'e}mat},
  {Garc{\'\i}a-Torres}, {Gosset}, {Halbwachs}, {Hambly}, {Harrison}, {Hauser},
  {Hestroffer}, {Hodgkin}, {Huckle}, {Hutton}, {Jasniewicz}, {Jordan},
  {Kontizas}, {Korn}, {Lanzafame}, {Manteiga}, {Moitinho}, {Muinonen},
  {Osinde}, {Pancino}, {Pauwels}, {Petit}, {Recio-Blanco}, {Robin}, {Sarro},
  {Siopis}, {Smith}, {Smith}, {Sozzetti}, {Thuillot}, {van Reeven}, {Viala},
  {Abbas}, {Abreu Aramburu}, {Accart}, {Aguado}, {Allan}, {Allasia},
  {Altavilla}, {{\'A}lvarez}, {Alves}, {Anderson}, {Andrei}, {Anglada Varela},
  {Antiche}, {Antoja}, {Ant{\'o}n}, {Arcay}, {Atzei}, {Ayache}, {Bach},
  {Baker}, {Balaguer-N{\'u}{\~n}ez}, {Barache}, {Barata}, {Barbier}, {Barblan},
  {Baroni}, {Barrado y Navascu{\'e}s}, {Barros}, {Barstow}, {Becciani},
  {Bellazzini}, {Bellei}, {Bello Garc{\'\i}a}, {Belokurov}, {Bendjoya},
  {Berihuete}, {Bianchi}, {Bienaym{\'e}}, {Billebaud}, {Blagorodnova},
  {Blanco-Cuaresma}, {Boch}, {Bombrun}, {Borrachero}, {Bouquillon}, {Bourda},
  {Bouy}, {Bragaglia}, {Breddels}, {Brouillet}, {Br{\"u}semeister},
  {Bucciarelli}, {Budnik}, {Burgess}, {Burgon}, {Burlacu}, {Busonero}, {Buzzi},
  {Caffau}, {Cambras}, {Campbell}, {Cancelliere}, {Cantat-Gaudin}, {Carlucci},
  {Carrasco}, {Castellani}, {Charlot}, {Charnas}, {Charvet}, {Chassat},
  {Chiavassa}, {Clotet}, {Cocozza}, {Collins}, {Collins}, {Costigan}, {Crifo},
  {Cross}, {Crosta}, {Crowley}, {Dafonte}, {Damerdji}, {Dapergolas}, {David},
  {David}, {De Cat}, {de Felice}, {de Laverny}, {De Luise}, {De March}, {de
  Martino}, {de Souza}, {Debosscher}, {del Pozo}, {Delbo}, {Delgado},
  {Delgado}, {di Marco}, {Di Matteo}, {Diakite}, {Distefano}, {Dolding}, {Dos
  Anjos}, {Drazinos}, {Dur{\'a}n}, {Dzigan}, {Ecale}, {Edvardsson}, {Enke},
  {Erdmann}, {Escolar}, {Espina}, {Evans}, {Eynard Bontemps}, {Fabre},
  {Fabrizio}, {Faigler}, {Falc{\~a}o}, {Farr{\`a}s Casas}, {Faye}, {Federici},
  {Fedorets}, {Fern{\'a}ndez-Hern{\'a}ndez}, {Fernique}, {Fienga}, {Figueras},
  {Filippi}, {Findeisen}, {Fonti}, {Fouesneau}, {Fraile}, {Fraser}, {Fuchs},
  {Furnell}, {Gai}, {Galleti}, {Galluccio}, {Garabato}, {Garc{\'\i}a-Sedano},
  {Gar{\'e}}, {Garofalo}, {Garralda}, {Gavras}, {Gerssen}, {Geyer}, {Gilmore},
  {Girona}, {Giuffrida}, {Gomes}, {Gonz{\'a}lez-Marcos},
  {Gonz{\'a}lez-N{\'u}{\~n}ez}, {Gonz{\'a}lez-Vidal}, {Granvik}, {Guerrier},
  {Guillout}, {Guiraud}, {G{\'u}rpide}, {Guti{\'e}rrez-S{\'a}nchez}, {Guy},
  {Haigron}, {Hatzidimitriou}, {Haywood}, {Heiter}, {Helmi}, {Hobbs},
  {Hofmann}, {Holl}, {Holland }, {Hunt}, {Hypki}, {Icardi}, {Irwin}, {Jevardat
  de Fombelle}, {Jofr{\'e}}, {Jonker}, {Jorissen}, {Julbe}, {Karampelas},
  {Kochoska}, {Kohley}, {Kolenberg}, {Kontizas}, {Koposov}, {Kordopatis},
  {Koubsky}, {Kowalczyk}, {Krone-Martins}, {Kudryashova}, {Kull}, {Bachchan},
  {Lacoste-Seris}, {Lanza}, {Lavigne}, {Le Poncin-Lafitte}, {Lebreton},
  {Lebzelter}, {Leccia}, {Leclerc}, {Lecoeur-Taibi}, {Lemaitre}, {Lenhardt},
  {Leroux}, {Liao}, {Licata}, {Lindstr{\o}m}, {Lister}, {Livanou}, {Lobel},
  {L{\"o}ffler}, {L{\'o}pez}, {Lopez-Lozano}, {Lorenz}, {Loureiro},
  {MacDonald}, {Magalh{\~a}es Fernandes}, {Managau}, {Mann}, {Mantelet},
  {Marchal}, {Marchant}, {Marconi}, {Marie}, {Marinoni}, {Marrese},
  {Marschalk{\'o}}, {Marshall}, {Mart{\'\i}n-Fleitas}, {Martino}, {Mary},
  {Matijevi{\v{c}}}, {Mazeh}, {McMillan}, {Messina}, {Mestre}, {Michalik},
  {Millar}, {Miranda}, {Molina}, {Molinaro}, {Molinaro}, {Moln{\'a}r},
  {Moniez}, {Montegriffo}, {Monteiro}, {Mor}, {Mora}, {Morbidelli}, {Morel},
  {Morgenthaler}, {Morley}, {Morris}, {Mulone}, {Muraveva}, {Musella},
  {Narbonne}, {Nelemans}, {Nicastro}, {Noval}, {Ord{\'e}novic},
  {Ordieres-Mer{\'e}}, {Osborne}, {Pagani}, {Pagano}, {Pailler}, {Palacin},
  {Palaversa}, {Parsons}, {Paulsen}, {Pecoraro}, {Pedrosa}, {Pentik{\"a}inen},
  {Pereira}, {Pichon}, {Piersimoni}, {Pineau}, {Plachy}, {Plum}, {Poujoulet},
  {Pr{\v{s}}a}, {Pulone}, {Ragaini}, {Rago}, {Rambaux}, {Ramos-Lerate},
  {Ranalli}, {Rauw}, {Read}, {Regibo}, {Renk}, {Reyl{\'e}}, {Ribeiro},
  {Rimoldini}, {Ripepi}, {Riva}, {Rixon}, {Roelens}, {Romero-G{\'o}mez},
  {Rowell}, {Royer}, {Rudolph}, {Ruiz-Dern}, {Sadowski}, {Sagrist{\`a}
  Sell{\'e}s}, {Sahlmann}, {Salgado}, {Salguero}, {Sarasso}, {Savietto},
  {Schnorhk}, {Schultheis}, {Sciacca}, {Segol}, {Segovia}, {Segransan},
  {Serpell}, {Shih}, {Smareglia}, {Smart}, {Smith}, {Solano}, {Solitro},
  {Sordo}, {Soria Nieto}, {Souchay}, {Spagna}, {Spoto}, {Stampa}, {Steele},
  {Steidelm{\"u}ller}, {Stephenson}, {Stoev}, {Suess}, {S{\"u}veges}, {Surdej},
  {Szabados}, {Szegedi-Elek}, {Tapiador}, {Taris}, {Tauran}, {Taylor},
  {Teixeira}, {Terrett}, {Tingley}, {Trager}, {Turon}, {Ulla}, {Utrilla},
  {Valentini}, {van Elteren}, {Van Hemelryck}, {van Leeuwen}, {Varadi},
  {Vecchiato}, {Veljanoski}, {Via}, {Vicente}, {Vogt}, {Voss}, {Votruba},
  {Voutsinas}, {Walmsley}, {Weiler}, {Weingrill}, {Werner}, {Wevers},
  {Whitehead}, {Wyrzykowski}, {Yoldas}, {{\v{Z}}erjal}, {Zucker}, {Zurbach},
  {Zwitter}, {Alecu}, {Allen}, {Allende Prieto}, {Amorim},
  {Anglada-Escud{\'e}}, {Arsenijevic}, {Azaz}, {Balm}, {Beck}, {Bernstein},
  {Bigot}, {Bijaoui}, {Blasco}, {Bonfigli}, {Bono}, {Boudreault}, {Bressan},
  {Brown}, {Brunet}, {Bunclark}, {Buonanno}, {Butkevich}, {Carret}, {Carrion},
  {Chemin}, {Ch{\'e}reau}, {Corcione}, {Darmigny}, {de Boer}, {de Teodoro}, {de
  Zeeuw}, {Delle Luche}, {Domingues}, {Dubath}, {Fodor}, {Fr{\'e}zouls},
  {Fries}, {Fustes}, {Fyfe}, {Gallardo}, {Gallegos}, {Gardiol}, {Gebran},
  {Gomboc}, {G{\'o}mez}, {Grux}, {Gueguen}, {Heyrovsky}, {Hoar}, {Iannicola},
  {Isasi Parache}, {Janotto}, {Joliet}, {Jonckheere}, {Keil}, {Kim},
  {Klagyivik}, {Klar}, {Knude}, {Kochukhov}, {Kolka}, {Kos}, {Kutka}, {Lainey},
  {LeBouquin}, {Liu}, {Loreggia}, {Makarov}, {Marseille}, {Martayan},
  {Martinez-Rubi}, {Massart}, {Meynadier}, {Mignot}, {Munari}, {Nguyen},
  {Nordlander}, {Ocvirk}, {O'Flaherty}, {Olias Sanz}, {Ortiz}, {Osorio},
  {Oszkiewicz}, {Ouzounis}, {Palmer}, {Park}, {Pasquato}, {Peltzer}, {Peralta},
  {P{\'e}turaud}, {Pieniluoma}, {Pigozzi}, {Poels}, {Prat}, {Prod'homme},
  {Raison}, {Rebordao}, {Risquez}, {Rocca-Volmerange}, {Rosen}, {Ruiz-Fuertes},
  {Russo}, {Sembay}, {Serraller Vizcaino}, {Short}, {Siebert}, {Silva},
  {Sinachopoulos}, {Slezak}, {Soffel}, {Sosnowska}, {Strai{\v{z}}ys}, {ter
  Linden}, {Terrell}, {Theil}, {Tiede}, {Troisi}, {Tsalmantza}, {Tur},
  {Vaccari}, {Vachier}, {Valles}, {Van Hamme}, {Veltz}, {Virtanen}, {Wallut},
  {Wichmann}, {Wilkinson}, {Ziaeepour}, \& {Zschocke}}]{Gaia2016}
{Gaia Collaboration}, {Prusti}, T., {de Bruijne}, J.~H.~J., {et~al.} 2016,
  \aap, 595, A1, \dodoi{10.1051/0004-6361/201629272}

\bibitem[{{Gaia Collaboration} {et~al.}(2018){Gaia Collaboration}, {Brown},
  {Vallenari}, {Prusti}, {de Bruijne}, {Babusiaux}, {Bailer-Jones}, {Biermann},
  {Evans}, {Eyer}, {Jansen}, {Jordi}, {Klioner}, {Lammers}, {Lindegren},
  {Luri}, {Mignard}, {Panem}, {Pourbaix}, {Randich}, {Sartoretti}, {Siddiqui},
  {Soubiran}, {van Leeuwen}, {Walton}, {Arenou}, {Bastian}, {Cropper},
  {Drimmel}, {Katz}, {Lattanzi}, {Bakker}, {Cacciari}, {Casta{\~n}eda},
  {Chaoul}, {Cheek}, {De Angeli}, {Fabricius}, {Guerra}, {Holl}, {Masana},
  {Messineo}, {Mowlavi}, {Nienartowicz}, {Panuzzo}, {Portell}, {Riello},
  {Seabroke}, {Tanga}, {Th{\'e}venin}, {Gracia-Abril}, {Comoretto},
  {Garcia-Reinaldos}, {Teyssier}, {Altmann}, {Andrae}, {Audard},
  {Bellas-Velidis}, {Benson}, {Berthier}, {Blomme}, {Burgess}, {Busso},
  {Carry}, {Cellino}, {Clementini}, {Clotet}, {Creevey}, {Davidson}, {De
  Ridder}, {Delchambre}, {Dell'Oro}, {Ducourant},
  {Fern{\'a}ndez-Hern{\'a}ndez}, {Fouesneau}, {Fr{\'e}mat}, {Galluccio},
  {Garc{\'\i}a-Torres}, {Gonz{\'a}lez-N{\'u}{\~n}ez}, {Gonz{\'a}lez-Vidal},
  {Gosset}, {Guy}, {Halbwachs}, {Hambly}, {Harrison}, {Hern{\'a}ndez},
  {Hestroffer}, {Hodgkin}, {Hutton}, {Jasniewicz}, {Jean-Antoine-Piccolo},
  {Jordan}, {Korn}, {Krone-Martins}, {Lanzafame}, {Lebzelter}, {L{\"o}ffler},
  {Manteiga}, {Marrese}, {Mart{\'\i}n-Fleitas}, {Moitinho}, {Mora}, {Muinonen},
  {Osinde}, {Pancino}, {Pauwels}, {Petit}, {Recio-Blanco}, {Richards},
  {Rimoldini}, {Robin}, {Sarro}, {Siopis}, {Smith}, {Sozzetti}, {S{\"u}veges},
  {Torra}, {van Reeven}, {Abbas}, {Abreu Aramburu}, {Accart}, {Aerts},
  {Altavilla}, {{\'A}lvarez}, {Alvarez}, {Alves}, {Anderson}, {Andrei},
  {Anglada Varela}, {Antiche}, {Antoja}, {Arcay}, {Astraatmadja}, {Bach},
  {Baker}, {Balaguer-N{\'u}{\~n}ez}, {Balm}, {Barache}, {Barata}, {Barbato},
  {Barblan}, {Barklem}, {Barrado}, {Barros}, {Barstow}, {Bartholom{\'e}
  Mu{\~n}oz}, {Bassilana}, {Becciani}, {Bellazzini}, {Berihuete}, {Bertone},
  {Bianchi}, {Bienaym{\'e}}, {Blanco-Cuaresma}, {Boch}, {Boeche}, {Bombrun},
  {Borrachero}, {Bossini}, {Bouquillon}, {Bourda}, {Bragaglia}, {Bramante},
  {Breddels}, {Bressan}, {Brouillet}, {Br{\"u}semeister}, {Brugaletta},
  {Bucciarelli}, {Burlacu}, {Busonero}, {Butkevich}, {Buzzi}, {Caffau},
  {Cancelliere}, {Cannizzaro}, {Cantat-Gaudin}, {Carballo}, {Carlucci},
  {Carrasco}, {Casamiquela}, {Castellani}, {Castro-Ginard}, {Charlot},
  {Chemin}, {Chiavassa}, {Cocozza}, {Costigan}, {Cowell}, {Crifo}, {Crosta},
  {Crowley}, {Cuypers}, {Dafonte}, {Damerdji}, {Dapergolas}, {David}, {David},
  {de Laverny}, {De Luise}, {De March}, {de Martino}, {de Souza}, {de Torres},
  {Debosscher}, {del Pozo}, {Delbo}, {Delgado}, {Delgado}, {Di Matteo},
  {Diakite}, {Diener}, {Distefano}, {Dolding}, {Drazinos}, {Dur{\'a}n},
  {Edvardsson}, {Enke}, {Eriksson}, {Esquej}, {Eynard Bontemps}, {Fabre},
  {Fabrizio}, {Faigler}, {Falc{\~a}o}, {Farr{\`a}s Casas}, {Federici},
  {Fedorets}, {Fernique}, {Figueras}, {Filippi}, {Findeisen}, {Fonti},
  {Fraile}, {Fraser}, {Fr{\'e}zouls}, {Gai}, {Galleti}, {Garabato},
  {Garc{\'\i}a-Sedano}, {Garofalo}, {Garralda}, {Gavel}, {Gavras}, {Gerssen},
  {Geyer}, {Giacobbe}, {Gilmore}, {Girona}, {Giuffrida}, {Glass}, {Gomes},
  {Granvik}, {Gueguen}, {Guerrier}, {Guiraud}, {Guti{\'e}rrez-S{\'a}nchez},
  {Haigron}, {Hatzidimitriou}, {Hauser}, {Haywood}, {Heiter}, {Helmi}, {Heu},
  {Hilger}, {Hobbs}, {Hofmann}, {Holland}, {Huckle}, {Hypki}, {Icardi},
  {Jan{\ss}en}, {Jevardat de Fombelle}, {Jonker}, {Juh{\'a}sz}, {Julbe},
  {Karampelas}, {Kewley}, {Klar}, {Kochoska}, {Kohley}, {Kolenberg},
  {Kontizas}, {Kontizas}, {Koposov}, {Kordopatis}, {Kostrzewa-Rutkowska},
  {Koubsky}, {Lambert}, {Lanza}, {Lasne}, {Lavigne}, {Le Fustec}, {Le
  Poncin-Lafitte}, {Lebreton}, {Leccia}, {Leclerc}, {Lecoeur-Taibi},
  {Lenhardt}, {Leroux}, {Liao}, {Licata}, {Lindstr{\o}m}, {Lister}, {Livanou},
  {Lobel}, {L{\'o}pez}, {Managau}, {Mann}, {Mantelet}, {Marchal}, {Marchant},
  {Marconi}, {Marinoni}, {Marschalk{\'o}}, {Marshall}, {Martino}, {Marton},
  {Mary}, {Massari}, {Matijevi{\v{c}}}, {Mazeh}, {McMillan}, {Messina},
  {Michalik}, {Millar}, {Molina}, {Molinaro}, {Moln{\'a}r}, {Montegriffo},
  {Mor}, {Morbidelli}, {Morel}, {Morris}, {Mulone}, {Muraveva}, {Musella},
  {Nelemans}, {Nicastro}, {Noval}, {O'Mullane}, {Ord{\'e}novic},
  {Ord{\'o}{\~n}ez-Blanco}, {Osborne}, {Pagani}, {Pagano}, {Pailler},
  {Palacin}, {Palaversa}, {Panahi}, {Pawlak}, {Piersimoni}, {Pineau}, {Plachy},
  {Plum}, {Poggio}, {Poujoulet}, {Pr{\v{s}}a}, {Pulone}, {Racero}, {Ragaini},
  {Rambaux}, {Ramos-Lerate}, {Regibo}, {Reyl{\'e}}, {Riclet}, {Ripepi}, {Riva},
  {Rivard}, {Rixon}, {Roegiers}, {Roelens}, {Romero-G{\'o}mez}, {Rowell},
  {Royer}, {Ruiz-Dern}, {Sadowski}, {Sagrist{\`a} Sell{\'e}s}, {Sahlmann},
  {Salgado}, {Salguero}, {Sanna}, {Santana-Ros}, {Sarasso}, {Savietto},
  {Schultheis}, {Sciacca}, {Segol}, {Segovia}, {S{\'e}gransan}, {Shih},
  {Siltala}, {Silva}, {Smart}, {Smith}, {Solano}, {Solitro}, {Sordo}, {Soria
  Nieto}, {Souchay}, {Spagna}, {Spoto}, {Stampa}, {Steele},
  {Steidelm{\"u}ller}, {Stephenson}, {Stoev}, {Suess}, {Surdej}, {Szabados},
  {Szegedi-Elek}, {Tapiador}, {Taris}, {Tauran}, {Taylor}, {Teixeira},
  {Terrett}, {Teyssand ier}, {Thuillot}, {Titarenko}, {Torra Clotet}, {Turon},
  {Ulla}, {Utrilla}, {Uzzi}, {Vaillant}, {Valentini}, {Valette}, {van Elteren},
  {Van Hemelryck}, {van Leeuwen}, {Vaschetto}, {Vecchiato}, {Veljanoski},
  {Viala}, {Vicente}, {Vogt}, {von Essen}, {Voss}, {Votruba}, {Voutsinas},
  {Walmsley}, {Weiler}, {Wertz}, {Wevers}, {Wyrzykowski}, {Yoldas},
  {{\v{Z}}erjal}, {Ziaeepour}, {Zorec}, {Zschocke}, {Zucker}, {Zurbach}, \&
  {Zwitter}}]{Gaia2018}
{Gaia Collaboration}, {Brown}, A.~G.~A., {Vallenari}, A., {et~al.} 2018, \aap,
  616, A1, \dodoi{10.1051/0004-6361/201833051}

\bibitem[{{Garc{\'\i}a P{\'e}rez} {et~al.}(2016){Garc{\'\i}a P{\'e}rez},
  {Allende Prieto}, {Holtzman}, {Shetrone}, {M{\'e}sz{\'a}ros}, {Bizyaev},
  {Carrera}, {Cunha}, {Garc{\'\i}a-Hern{\'a}ndez}, {Johnson}, {Majewski},
  {Nidever}, {Schiavon}, {Shane}, {Smith}, {Sobeck}, {Troup}, {Zamora},
  {Weinberg}, {Bovy}, {Eisenstein}, {Feuillet}, {Frinchaboy}, {Hayden},
  {Hearty}, {Nguyen}, {O'Connell}, {Pinsonneault}, {Wilson}, \&
  {Zasowski}}]{GarciaPerez}
{Garc{\'\i}a P{\'e}rez}, A.~E., {Allende Prieto}, C., {Holtzman}, J.~A.,
  {et~al.} 2016, \aj, 151, 144, \dodoi{10.3847/0004-6256/151/6/144}

\bibitem[{{Gilmore} {et~al.}(2012){Gilmore}, {Randich}, {Asplund}, {Binney},
  {Bonifacio}, {Drew}, {Feltzing}, {Ferguson}, {Jeffries}, {Micela},
  {Negueruela}, {Prusti}, {Rix}, {Vallenari}, {Alfaro}, {Allende-Prieto},
  {Babusiaux}, {Bensby}, {Blomme}, {Bragaglia}, {Flaccomio}, {Fran{\c{c}}ois},
  {Irwin}, {Koposov}, {Korn}, {Lanzafame}, {Pancino}, {Paunzen},
  {Recio-Blanco}, {Sacco}, {Smiljanic}, {Van Eck}, {Walton}, {Aden}, {Aerts},
  {Affer}, {Alcala}, {Altavilla}, {Alves}, {Antoja}, {Arenou}, {Argiroffi},
  {Asensio Ramos}, {Bailer-Jones}, {Balaguer-Nunez}, {Bayo}, {Barbuy},
  {Barisevicius}, {Barrado y Navascues}, {Battistini}, {Bellas Velidis},
  {Bellazzini}, {Belokurov}, {Bergemann}, {Bertelli}, {Biazzo}, {Bienayme},
  {Bland-Hawthorn}, {Boeche}, {Bonito}, {Boudreault}, {Bouvier}, {Brandao},
  {Brown}, {de Bruijne}, {Burleigh}, {Caballero}, {Caffau}, {Calura},
  {Capuzzo-Dolcetta}, {Caramazza}, {Carraro}, {Casagrande}, {Casewell},
  {Chapman}, {Chiappini}, {Chorniy}, {Christlieb}, {Cignoni}, {Cocozza},
  {Colless}, {Collet}, {Collins}, {Correnti}, {Covino}, {Crnojevic}, {Cropper},
  {Cunha}, {Damiani}, {David}, {Delgado}, {Duffau}, {Edvardsson}, {Eldridge},
  {Enke}, {Eriksson}, {Evans}, {Eyer}, {Famaey}, {Fellhauer}, {Ferreras},
  {Figueras}, {Fiorentino}, {Flynn}, {Folha}, {Franciosini}, {Frasca},
  {Freeman}, {Fremat}, {Friel}, {Gaensicke}, {Gameiro}, {Garzon}, {Geier},
  {Geisler}, {Gerhard}, {Gibson}, {Gomboc}, {Gomez}, {Gonzalez-Fernandez},
  {Gonzalez Hernandez}, {Gosset}, {Grebel}, {Greimel}, {Groenewegen},
  {Grundahl}, {Guarcello}, {Gustafsson}, {Hadrava}, {Hatzidimitriou}, {Hambly},
  {Hammersley}, {Hansen}, {Haywood}, {Heber}, {Heiter}, {Held}, {Helmi},
  {Hensler}, {Herrero}, {Hill}, {Hodgkin}, {Huelamo}, {Huxor}, {Ibata},
  {Jackson}, {de Jong}, {Jonker}, {Jordan}, {Jordi}, {Jorissen}, {Katz},
  {Kawata}, {Keller}, {Kharchenko}, {Klement}, {Klutsch}, {Knude}, {Koch},
  {Kochukhov}, {Kontizas}, {Koubsky}, {Lallement}, {de Laverny}, {van Leeuwen},
  {Lemasle}, {Lewis}, {Lind}, {Lindstrom}, {Lobel}, {Lopez Santiago}, {Lucas},
  {Ludwig}, {Lueftinger}, {Magrini}, {Maiz Apellaniz}, {Maldonado}, {Marconi},
  {Marino}, {Martayan}, {Martinez-Valpuesta}, {Matijevic}, {McMahon},
  {Messina}, {Meyer}, {Miglio}, {Mikolaitis}, {Minchev}, {Minniti}, {Moitinho},
  {Momany}, {Monaco}, {Montalto}, {Monteiro}, {Monier}, {Montes}, {Mora},
  {Moraux}, {Morel}, {Mowlavi}, {Mucciarelli}, {Munari}, {Napiwotzki},
  {Nardetto}, {Naylor}, {Naze}, {Nelemans}, {Okamoto}, {Ortolani}, {Pace},
  {Palla}, {Palous}, {Parker}, {Penarrubia}, {Pillitteri}, {Piotto}, {Posbic},
  {Prisinzano}, {Puzeras}, {Quirrenbach}, {Ragaini}, {Read}, {Read}, {Reyle},
  {De Ridder}, {Robichon}, {Robin}, {Roeser}, {Romano}, {Royer}, {Ruchti},
  {Ruzicka}, {Ryan}, {Ryde}, {Santos}, {Sanz Forcada}, {Sarro Baro},
  {Sbordone}, {Schilbach}, {Schmeja}, {Schnurr}, {Schoenrich}, {Scholz},
  {Seabroke}, {Sharma}, {De Silva}, {Smith}, {Solano}, {Sordo}, {Soubiran},
  {Sousa}, {Spagna}, {Steffen}, {Steinmetz}, {Stelzer}, {Stempels},
  {Tabernero}, {Tautvaisiene}, {Thevenin}, {Torra}, {Tosi}, {Tolstoy}, {Turon},
  {Walker}, {Wambsganss}, {Worley}, {Venn}, {Vink}, {Wyse}, {Zaggia},
  {Zeilinger}, {Zoccali}, {Zorec}, {Zucker}, {Zwitter}, \& {Gaia-ESO Survey
  Team}}]{gilmore+2012}
{Gilmore}, G., {Randich}, S., {Asplund}, M., {et~al.} 2012, The Messenger, 147,
  25

\bibitem[{{Holmberg} {et~al.}(2009){Holmberg}, {Nordstr{\"o}m}, \&
  {Andersen}}]{holmberg_nordstrom_andersen2009}
{Holmberg}, J., {Nordstr{\"o}m}, B., \& {Andersen}, J. 2009, \aap, 501, 941

\bibitem[{{Jofr{\'e}} {et~al.}(2014){Jofr{\'e}}, {Heiter}, {Soubiran},
  {Blanco-Cuaresma}, {Worley}, {Pancino}, {Cantat-Gaudin}, {Magrini},
  {Bergemann}, {Gonz{\'a}lez Hern{\'a}ndez}, {Hill}, {Lardo}, {de Laverny},
  {Lind}, {Masseron}, {Montes}, {Mucciarelli}, {Nordlander}, {Recio Blanco},
  {Sobeck}, {Sordo}, {Sousa}, {Tabernero}, {Vallenari}, \& {Van
  Eck}}]{jofre2014}
{Jofr{\'e}}, P., {Heiter}, U., {Soubiran}, C., {et~al.} 2014, \aap, 564, A133,
  \dodoi{10.1051/0004-6361/201322440}

\bibitem[{{Johnson} \& {Soderblom}(1987)}]{JohnsonSoderblom1987}
{Johnson}, D. R.~H., \& {Soderblom}, D.~R. 1987, \aj, 93, 864,
  \dodoi{10.1086/114370}

\bibitem[{{Johnson} \& {Li}(2012)}]{johnson2012}
{Johnson}, J.~L., \& {Li}, H. 2012, \apj, 751, 81,
  \dodoi{10.1088/0004-637X/751/2/81}

\bibitem[{{J{\"o}nsson} {et~al.}(2020){J{\"o}nsson}, {Holtzman}, {Allende
  Prieto}, {Cunha}, {Garc{\'\i}a-Hern{\'a}ndez}, {Hasselquist}, {Masseron},
  {Osorio}, {Shetrone}, {Smith}, {Stringfellow}, {Bizyaev}, {Edvardsson},
  {Majewski}, {M{\'e}sz{\'a}ros}, {Souto}, {Zamora}, {Beaton}, {Bovy}, {Donor},
  {Pinsonneault}, {Poovelil}, \& {Sobeck}}]{APOGEEDR16}
{J{\"o}nsson}, H., {Holtzman}, J.~A., {Allende Prieto}, C., {et~al.} 2020, \aj,
  160, 120, \dodoi{10.3847/1538-3881/aba592}

\bibitem[{Kramida {et~al.}(2020)Kramida, {Yu.~Ralchenko}, Reader, \& {and NIST
  ASD Team}}]{NIST_ASD}
Kramida, A., {Yu.~Ralchenko}, Reader, J., \& {and NIST ASD Team}. 2020, {NIST
  Atomic Spectra Database (ver. 5.8), [Online]. Available:
  {\tt{https://physics.nist.gov/asd}} [2021, April 19]. National Institute of
  Standards and Technology, Gaithersburg, MD.}

\bibitem[{{Kunder} {et~al.}(2017){Kunder}, {Kordopatis}, {Steinmetz},
  {Zwitter}, {McMillan}, {Casagrande}, {Enke}, {Wojno}, {Valentini},
  {Chiappini}, {Matijevi{\v{c}}}, {Siviero}, {de Laverny}, {Recio-Blanco},
  {Bijaoui}, {Wyse}, {Binney}, {Grebel}, {Helmi}, {Jofre}, {Antoja}, {Gilmore},
  {Siebert}, {Famaey}, {Bienaym{\'e}}, {Gibson}, {Freeman}, {Navarro},
  {Munari}, {Seabroke}, {Anguiano}, {{\v{Z}}erjal}, {Minchev}, {Reid},
  {Bland-Hawthorn}, {Kos}, {Sharma}, {Watson}, {Parker}, {Scholz}, {Burton},
  {Cass}, {Hartley}, {Fiegert}, {Stupar}, {Ritter}, {Hawkins}, {Gerhard},
  {Chaplin}, {Davies}, {Elsworth}, {Lund}, {Miglio}, \& {Mosser}}]{kunder+2017}
{Kunder}, A., {Kordopatis}, G., {Steinmetz}, M., {et~al.} 2017, \aj, 153, 75

\bibitem[{{Luo} {et~al.}(2015){Luo}, {Zhao}, {Zhao}, {Deng}, {Liu}, {Jing},
  {Wang}, {Zhang}, {Shi}, {Cui}, {Chu}, {Li}, {Bai}, {Wu}, {Cai}, {Cao}, {Cao},
  {Carlin}, {Chen}, {Chen}, {Chen}, {Chen}, {Chen}, {Chen}, {Chen},
  {Christlieb}, {Chu}, {Cui}, {Dong}, {Du}, {Fan}, {Feng}, {Fu}, {Gao}, {Gong},
  {Gu}, {Guo}, {Han}, {He}, {Hou}, {Hou}, {Hou}, {Hu}, {Hu}, {Hu}, {Huo},
  {Jia}, {Jiang}, {Jiang}, {Jiang}, {Jin}, {Kong}, {Kong}, {Lei}, {Li}, {Li},
  {Li}, {Li}, {Li}, {Li}, {Li}, {Li}, {Li}, {Li}, {Li}, {Li}, {Liang}, {Lin},
  {Liu}, {Liu}, {Liu}, {Liu}, {Lu}, {Luo}, {Mao}, {Newberg}, {Ni}, {Qi}, {Qi},
  {Shen}, {Shi}, {Song}, {Song}, {Su}, {Su}, {Tang}, {Tao}, {Tian}, {Wang},
  {Wang}, {Wang}, {Wang}, {Wang}, {Wang}, {Wang}, {Wang}, {Wang}, {Wang},
  {Wang}, {Wang}, {Wang}, {Wang}, {Wang}, {Wang}, {Wang}, {Wang}, {Wang},
  {Wang}, {Wei}, {Wei}, {Wu}, {Wu}, {Wu}, {Wu}, {Xing}, {Xu}, {Xu}, {Xu},
  {Yan}, {Yang}, {Yang}, {Yang}, {Yang}, {Yao}, {Yu}, {Yuan}, {Yuan}, {Yuan},
  {Yuan}, {Zhai}, {Zhang}, {Zhang}, {Zhang}, {Zhang}, {Zhang}, {Zhang},
  {Zhang}, {Zhang}, {Zhao}, {Zhou}, {Zhou}, {Zhu}, {Zhu}, {Zou}, \&
  {Zuo}}]{luo+2015}
{Luo}, A.~L., {Zhao}, Y.-H., {Zhao}, G., {et~al.} 2015, Research in Astronomy
  and Astrophysics, 15, 1095

\bibitem[{{M.~Kovalev} {et~al.}(2018){M.~Kovalev}, {S.~Brinkmann},
  {M.~Bergemann}, \& {MPIA IT-department}}]{NLTE_MPIA}
{M.~Kovalev}, {S.~Brinkmann}, {M.~Bergemann}, \& {MPIA IT-department}. 2018,
  {NLTE MPIA web server, [Online]. Available: {{http://nlte.mpia.de}} Max
  Planck Institute for Astronomy, Heidelberg.}

\bibitem[{{Mack} {et~al.}(2018){Mack}, {Strassmeier}, {Ilyin}, {Schuler},
  {Spada}, \& {Barnes}}]{Mack+2018}
{Mack}, C.~E., {Strassmeier}, K.~G., {Ilyin}, I., {et~al.} 2018, \aap, 612,
  A46, \dodoi{10.1051/0004-6361/201731634}

\bibitem[{{Majewski} {et~al.}(2017){Majewski}, {Schiavon}, {Frinchaboy},
  {Allende Prieto}, {Barkhouser}, {Bizyaev}, {Blank}, {Brunner}, {Burton},
  {Carrera}, {Chojnowski}, {Cunha}, {Epstein}, {Fitzgerald}, {Garc{\'\i}a
  P{\'e}rez}, {Hearty}, {Henderson}, {Holtzman}, {Johnson}, {Lam}, {Lawler},
  {Maseman}, {M{\'e}sz{\'a}ros}, {Nelson}, {Nguyen}, {Nidever}, {Pinsonneault},
  {Shetrone}, {Smee}, {Smith}, {Stolberg}, {Skrutskie}, {Walker}, {Wilson},
  {Zasowski}, {Anders}, {Basu}, {Beland}, {Blanton}, {Bovy}, {Brownstein},
  {Carlberg}, {Chaplin}, {Chiappini}, {Eisenstein}, {Elsworth}, {Feuillet},
  {Fleming}, {Galbraith-Frew}, {Garc{\'\i}a}, {Garc{\'\i}a-Hern{\'a}ndez},
  {Gillespie}, {Girardi}, {Gunn}, {Hasselquist}, {Hayden}, {Hekker}, {Ivans},
  {Kinemuchi}, {Klaene}, {Mahadevan}, {Mathur}, {Mosser}, {Muna}, {Munn},
  {Nichol}, {O'Connell}, {Parejko}, {Robin}, {Rocha-Pinto}, {Schultheis},
  {Serenelli}, {Shane}, {Silva Aguirre}, {Sobeck}, {Thompson}, {Troup},
  {Weinberg}, \& {Zamora}}]{Majewski}
{Majewski}, S.~R., {Schiavon}, R.~P., {Frinchaboy}, P.~M., {et~al.} 2017, \aj,
  154, 94, \dodoi{10.3847/1538-3881/aa784d}

\bibitem[{{M{\'e}sz{\'a}ros} {et~al.}(2012){M{\'e}sz{\'a}ros}, {Allende
  Prieto}, {Edvardsson}, {Castelli}, {Garc{\'\i}a P{\'e}rez}, {Gustafsson},
  {Majewski}, {Plez}, {Schiavon}, {Shetrone}, \& {de Vicente}}]{ATLAS9}
{M{\'e}sz{\'a}ros}, S., {Allende Prieto}, C., {Edvardsson}, B., {et~al.} 2012,
  \aj, 144, 120, \dodoi{10.1088/0004-6256/144/4/120}

\bibitem[{{Muirhead} {et~al.}(2018){Muirhead}, {Dressing}, {Mann},
  {Rojas-Ayala}, {L{\'e}pine}, {Paegert}, {De Lee}, \&
  {Oelkers}}]{muirhead+2018}
{Muirhead}, P.~S., {Dressing}, C.~D., {Mann}, A.~W., {et~al.} 2018, \aj, 155,
  180

\bibitem[{{Nidever} {et~al.}(2015){Nidever}, {Holtzman}, {Allende Prieto},
  {Beland}, {Bender}, {Bizyaev}, {Burton}, {Desphande}, {Fleming}, {Garc{\'\i}a
  P{\'e}rez}, {Hearty}, {Majewski}, {M{\'e}sz{\'a}ros}, {Muna}, {Nguyen},
  {Schiavon}, {Shetrone}, {Skrutskie}, {Sobeck}, \& {Wilson}}]{Nidever}
{Nidever}, D.~L., {Holtzman}, J.~A., {Allende Prieto}, C., {et~al.} 2015, \aj,
  150, 173, \dodoi{10.1088/0004-6256/150/6/173}

\bibitem[{{Nissen} \& {Schuster}(2010)}]{nissen_schuster2010}
{Nissen}, P.~E., \& {Schuster}, W.~J. 2010, \aap, 511, L10

\bibitem[{{Palme} {et~al.}(2014){Palme}, {Lodders}, \& {Jones}}]{Palme+2014}
{Palme}, H., {Lodders}, K., \& {Jones}, A. 2014, {Solar System Abundances of
  the Elements}, ed. A.~M. {Davis}, Vol.~2, 15--36

\bibitem[{{Petigura} {et~al.}(2018){Petigura}, {Marcy}, {Winn}, {Weiss},
  {Fulton}, {Howard}, {Sinukoff}, {Isaacson}, {Morton}, \&
  {Johnson}}]{Petigura2018}
{Petigura}, E.~A., {Marcy}, G.~W., {Winn}, J.~N., {et~al.} 2018, \aj, 155, 89,
  \dodoi{10.3847/1538-3881/aaa54c}

\bibitem[{{Posti} {et~al.}(2018){Posti}, {Helmi}, {Veljanoski}, \&
  {Breddels}}]{posti+2018}
{Posti}, L., {Helmi}, A., {Veljanoski}, J., \& {Breddels}, M.~A. 2018, \aap,
  615, A70

\bibitem[{Price-Whelan {et~al.}(2018)Price-Whelan, Sip{\H{o}}cz, G{\"u}nther,
  Lim, Crawford, Conseil, Shupe, Craig, Dencheva, Ginsburg, {et~al.}}]{AstroPy}
Price-Whelan, A.~M., Sip{\H{o}}cz, B., G{\"u}nther, H., {et~al.} 2018, The
  Astronomical Journal, 156, 123

\bibitem[{{Reddy} {et~al.}(2006){Reddy}, {Lambert}, \& {Allende
  Prieto}}]{Reddy+2006}
{Reddy}, B.~E., {Lambert}, D.~L., \& {Allende Prieto}, C. 2006, \mnras, 367,
  1329, \dodoi{10.1111/j.1365-2966.2006.10148.x}

\bibitem[{{Ricker} {et~al.}(2015){Ricker}, {Winn}, {Vanderspek}, {Latham},
  {Bakos}, {Bean}, {Berta-Thompson}, {Brown}, {Buchhave}, {Butler}, {Butler},
  {Chaplin}, {Charbonneau}, {Christensen-Dalsgaard}, {Clampin}, {Deming},
  {Doty}, {De Lee}, {Dressing}, {Dunham}, {Endl}, {Fressin}, {Ge}, {Henning},
  {Holman}, {Howard}, {Ida}, {Jenkins}, {Jernigan}, {Johnson}, {Kaltenegger},
  {Kawai}, {Kjeldsen}, {Laughlin}, {Levine}, {Lin}, {Lissauer}, {MacQueen},
  {Marcy}, {McCullough}, {Morton}, {Narita}, {Paegert}, {Palle}, {Pepe},
  {Pepper}, {Quirrenbach}, {Rinehart}, {Sasselov}, {Sato}, {Seager},
  {Sozzetti}, {Stassun}, {Sullivan}, {Szentgyorgyi}, {Torres}, {Udry}, \&
  {Villasenor}}]{Ricker2015}
{Ricker}, G.~R., {Winn}, J.~N., {Vanderspek}, R., {et~al.} 2015, Journal of
  Astronomical Telescopes, Instruments, and Systems, 1, 014003,
  \dodoi{10.1117/1.JATIS.1.1.014003}

\bibitem[{{Robin} {et~al.}(2003){Robin}, {Reyl{\'e}}, {Derri{\`e}re}, \&
  {Picaud}}]{robin+2003}
{Robin}, A.~C., {Reyl{\'e}}, C., {Derri{\`e}re}, S., \& {Picaud}, S. 2003,
  \aap, 409, 523

\bibitem[{{Sandage} \& {Fouts}(1987)}]{sandage1987}
{Sandage}, A., \& {Fouts}, G. 1987, \aj, 93, 592, \dodoi{10.1086/114341}

\bibitem[{{Schuster} {et~al.}(2012){Schuster}, {Moreno}, {Nissen}, \&
  {Pichardo}}]{schuster+2012}
{Schuster}, W.~J., {Moreno}, E., {Nissen}, P.~E., \& {Pichardo}, B. 2012, \aap,
  538, A21

\bibitem[{{Sharma} {et~al.}(2011){Sharma}, {Bland-Hawthorn}, {Johnston}, \&
  {Binney}}]{sharma+2011}
{Sharma}, S., {Bland-Hawthorn}, J., {Johnston}, K.~V., \& {Binney}, J. 2011,
  \apj, 730, 3, \dodoi{10.1088/0004-637X/730/1/3}

\bibitem[{{Sharma} {et~al.}(2018){Sharma}, {Stello}, {Buder}, {Kos},
  {Bland-Hawthorn}, {Asplund}, {Duong}, {Lin}, {Lind}, {Ness}, {Huber},
  {Zwitter}, {Traven}, {Hon}, {Kafle}, {Khanna}, {Saddon}, {Anguiano}, {Casey},
  {Freeman}, {Martell}, {De Silva}, {Simpson}, {Wittenmyer}, \&
  {Zucker}}]{sharma+2018}
{Sharma}, S., {Stello}, D., {Buder}, S., {et~al.} 2018, \mnras, 473, 2004

\bibitem[{{Skrutskie} {et~al.}(2006){Skrutskie}, {Cutri}, {Stiening},
  {Weinberg}, {Schneider}, {Carpenter}, {Beichman}, {Capps}, {Chester},
  {Elias}, {Huchra}, {Liebert}, {Lonsdale}, {Monet}, {Price}, {Seitzer},
  {Jarrett}, {Kirkpatrick}, {Gizis}, {Howard}, {Evans}, {Fowler}, {Fullmer},
  {Hurt}, {Light}, {Kopan}, {Marsh}, {McCallon}, {Tam}, {Van Dyk}, \&
  {Wheelock}}]{2MASS}
{Skrutskie}, M.~F., {Cutri}, R.~M., {Stiening}, R., {et~al.} 2006, \aj, 131,
  1163, \dodoi{10.1086/498708}

\bibitem[{{Smith} {et~al.}(2009){Smith}, {Evans}, {Belokurov}, {Hewett},
  {Bramich}, {Gilmore}, {Irwin}, {Vidrih}, \& {Zucker}}]{smith+2009}
{Smith}, M.~C., {Evans}, N.~W., {Belokurov}, V., {et~al.} 2009, \mnras, 399,
  1223

\bibitem[{{Sneden}(1973)}]{Sneden1973}
{Sneden}, C. 1973, \apj, 184, 839, \dodoi{10.1086/152374}

\bibitem[{{Soubiran} {et~al.}(2016){Soubiran}, {Le Campion}, {Brouillet}, \&
  {Chemin}}]{soubiran+2016}
{Soubiran}, C., {Le Campion}, J.-F., {Brouillet}, N., \& {Chemin}, L. 2016,
  \aap, 591, A118

\bibitem[{{Sozzetti} {et~al.}(2006){Sozzetti}, {Torres}, {Latham}, {Carney},
  {Stefanik}, {Boss}, {Laird}, \& {Korzennik}}]{Sozzetti2006}
{Sozzetti}, A., {Torres}, G., {Latham}, D.~W., {et~al.} 2006, ApJ, 649, 428,
  \dodoi{10.1086/506267}

\bibitem[{{Sozzetti} {et~al.}(2009){Sozzetti}, {Torres}, {Latham}, {Stefanik},
  {Korzennik}, {Boss}, {Carney}, \& {Laird}}]{Sozzetti+2009}
---. 2009, \apj, 697, 544, \dodoi{10.1088/0004-637X/697/1/544}

\bibitem[{{Stassun} {et~al.}(2018){Stassun}, {Oelkers}, {Pepper}, {Paegert},
  {De Lee}, {Torres}, {Latham}, {Charpinet}, {Dressing}, {Huber}, {Kane},
  {L{\'e}pine}, {Mann}, {Muirhead}, {Rojas-Ayala}, {Silvotti}, {Fleming},
  {Levine}, \& {Plavchan}}]{stassun+2018}
{Stassun}, K.~G., {Oelkers}, R.~J., {Pepper}, J., {et~al.} 2018, \aj, 156, 102

\bibitem[{{Stassun} {et~al.}(2019){Stassun}, {Oelkers}, {Paegert}, {Torres},
  {Pepper}, {De Lee}, {Collins}, {Latham}, {Muirhead}, {Chittidi},
  {Rojas-Ayala}, {Fleming}, {Rose}, {Tenenbaum}, {Ting}, {Kane}, {Barclay},
  {Bean}, {Brassuer}, {Charbonneau}, {Ge}, {Lissauer}, {Mann}, {McLean},
  {Mullally}, {Narita}, {Plavchan}, {Ricker}, {Sasselov}, {Seager}, {Sharma},
  {Shiao}, {Sozzetti}, {Stello}, {Vanderspek}, {Wallace}, \&
  {Winn}}]{stassun+2019}
{Stassun}, K.~G., {Oelkers}, R.~J., {Paegert}, M., {et~al.} 2019, \aj, 158, 138

\bibitem[{{Strassmeier} {et~al.}(2018){Strassmeier}, {Ilyin}, \&
  {Steffen}}]{Strassmeier2018}
{Strassmeier}, K.~G., {Ilyin}, I., \& {Steffen}, M. 2018, \aap, 612, A44,
  \dodoi{10.1051/0004-6361/201731631}

\bibitem[{{Strassmeier} {et~al.}(2015){Strassmeier}, {Ilyin}, {J{\"a}rvinen},
  {Weber}, {Woche}, {Barnes}, {Bauer}, {Beckert}, {Bittner}, {Bredthauer},
  {Carroll}, {Denker}, {Dionies}, {DiVarano}, {D{\"o}scher}, {Fechner},
  {Feuerstein}, {Granzer}, {Hahn}, {Harnisch}, {Hofmann}, {Lesser}, {Paschke},
  {Pankratow}, {Plank}, {Pl{\"u}schke}, {Popow}, \&
  {Sablowski}}]{Strassmeier2015}
{Strassmeier}, K.~G., {Ilyin}, I., {J{\"a}rvinen}, A., {et~al.} 2015,
  Astronomische Nachrichten, 336, 324, \dodoi{10.1002/asna.201512172}

\bibitem[{{Th{\'e}venin} \& {Idiart}(1999)}]{Thevenin1999}
{Th{\'e}venin}, F., \& {Idiart}, T.~P. 1999, \apj, 521, 753,
  \dodoi{10.1086/307578}

\bibitem[{Virtanen {et~al.}(2020)Virtanen, Gommers, Oliphant, Haberland, Reddy,
  Cournapeau, Burovski, Peterson, Weckesser, Bright, {van der Walt}, Brett,
  Wilson, Millman, Mayorov, Nelson, Jones, Kern, Larson, Carey, Polat, Feng,
  Moore, {VanderPlas}, Laxalde, Perktold, Cimrman, Henriksen, Quintero, Harris,
  Archibald, Ribeiro, Pedregosa, {van Mulbregt}, \& {SciPy 1.0
  Contributors}}]{Scipy}
Virtanen, P., Gommers, R., Oliphant, T.~E., {et~al.} 2020, Nature Methods, 17,
  261, \dodoi{10.1038/s41592-019-0686-2}

\bibitem[{{Wang} \& {Fischer}(2015)}]{Wang2015}
{Wang}, J., \& {Fischer}, D.~A. 2015, AJ, 149, 14,
  \dodoi{10.1088/0004-6256/149/1/14}

\bibitem[{{Wilson} {et~al.}(2018){Wilson}, {Teske}, {Majewski}, {Cunha},
  {Smith}, {Souto}, {Bender}, {Mahadevan}, {Troup}, {Allende Prieto},
  {Stassun}, {Skrutskie}, {Almeida}, {Garc{\'{\i}}a-Hern{\'a}ndez}, {Zamora},
  \& {Brinkmann}}]{Wilson2018}
{Wilson}, R.~F., {Teske}, J., {Majewski}, S.~R., {et~al.} 2018, AJ, 155, 68,
  \dodoi{10.3847/1538-3881/aa9f27}

\bibitem[{{Wright} {et~al.}(2010){Wright}, {Eisenhardt}, {Mainzer}, {Ressler},
  {Cutri}, {Jarrett}, {Kirkpatrick}, {Padgett}, {McMillan}, {Skrutskie},
  {Stanford}, {Cohen}, {Walker}, {Mather}, {Leisawitz}, {Gautier}, {McLean},
  {Benford}, {Lonsdale}, {Blain}, {Mendez}, {Irace}, {Duval}, {Liu}, {Royer},
  {Heinrichsen}, {Howard}, {Shannon}, {Kendall}, {Walsh}, {Larsen}, {Cardon},
  {Schick}, {Schwalm}, {Abid}, {Fabinsky}, {Naes}, \& {Tsai}}]{WISE}
{Wright}, E.~L., {Eisenhardt}, P. R.~M., {Mainzer}, A.~K., {et~al.} 2010, \aj,
  140, 1868, \dodoi{10.1088/0004-6256/140/6/1868}

\end{thebibliography}



\begin{longrotatetable}
\begin{deluxetable}{|cccc ccccccccccccc|}
\tablecaption{Table of PEPSI equivalent widths\label{table:EQWS}}
\tabletypesize{\footnotesize}
\tablenum{3}
\tablehead{
\hline
\multicolumn{17}{c}{} \cr
\multicolumn{4}{c}{\textbf{Line Information}} &
\multicolumn{13}{c}{\textbf{Equivalent Width for Given Star [m\AA]}} \cr
\hline
\colhead{} &
\colhead{} &
\colhead{} &
\colhead{} &
\colhead{BD+18} &
\colhead{BD+20} &
\colhead{BD+20} &
\colhead{BD+25} &
\colhead{BD+34} &
\colhead{BD+36} &
\colhead{BD+42} &
\colhead{BD+51} &
\colhead{BD+75} &
\colhead{HD} &
\colhead{HD} &
\colhead{HD} &
\colhead{HD} \cr
\colhead{Element} &
\colhead{$\lambda$ [\AA]} &
\colhead{EP [eV]} &
\colhead{log($gf$)}&
\colhead{3423} &
\colhead{2594} &
\colhead{3603} &
\colhead{1981} &
\colhead{2476} &
\colhead{2165} &
\colhead{2667} &
\colhead{1696} &
\colhead{839} &
\colhead{64090} &
\colhead{108177} &
\colhead{160693} &
\colhead{194598}
}

\startdata
Fe I & 4745.8 & 3.65 & -1.27 & ...  & 23.91  &  ...  &  ...  &  ...  &  ...  &  ...  &  ...  &  ...  &  ...  & 5.81  &  ...  &  ...  \\
Fe I & 4771.7 & 2.2 & -3.23 &11.71  &  ...  &  ...  &  ...  &  ...  &  ...  &  ...  &  ...  & 11.4  &  ...  &  ...  & 25.29  &  ...  \\
Fe I & 4772.83 & 3.02 & -2.19 &28.42  &  ...  &  ...  &  ...  &  ...  &  ...  & 14.16  &  ...  &  ...  &  ...  & 8.71  &  ...  &  ...  \\
Fe I & 4779.44 & 3.41 & -2.02 &6.56  & 7.01  &  ...  &  ...  &  ...  &  ...  &  ...  &  ...  &  ...  &  ...  &  ...  &  ...  &  ...  \\
Fe I & 4786.81 & 3.02 & -1.61 & ...  & 30.34  &  ...  &  ...  &  ...  & 9.26  & 15.46  &  ...  &  ...  & 23.91  & 7.5  & 54.96  &  ...  \\
Fe I & 4788.76 & 3.24 & -1.76 & ...  &  ...  &  ...  &  ...  &  ...  & 7.71  & 7.65  &  ...  &  ...  & 9.5  & 5.04  & 43.47  &  ...  \\
Fe I & 4789.65 & 3.55 & -0.96 &37.14  &  ...  &  ...  & 8.59  & 7.03  & 14.4  &  ...  &  ...  &  ...  & 26.11  & 11.05  & 60.75  & 28.97  \\
Fe I & 4791.25 & 3.27 & -2.44 & ...  & 9.4  &  ...  &  ...  &  ...  &  ...  &  ...  &  ...  &  ...  &  ...  &  ...  &  ...  &  ...  \\
Fe I & 4798.26 & 4.19 & -1.17 & ...  &  ...  &  ...  &  ...  &  ...  &  ...  &  ...  &  ...  &  ...  & 6.76  & 7.43  &  ...  &  ...  \\
Fe I & 4799.41 & 3.64 & -2.19 & ...  & 6.36  &  ...  &  ...  &  ...  &  ...  &  ...  &  ...  &  ...  &  ...  &  ...  &  ...  &  ...  \\
Fe I & 4800.65 & 4.14 & -1.03 &14.06  &  ...  &  ...  &  ...  &  ...  &  ...  &  ...  &  ...  &  ...  &  ...  &  ...  &  ...  &  ...  \\
Fe I & 4807.71 & 3.37 & -2.15 &6.22  &  ...  &  ...  &  ...  &  ...  &  ...  &  ...  &  ...  &  ...  &  ...  &  ...  &  ...  &  ...  \\
Fe I & 4817.78 & 2.22 & -3.44 &6.27  & 6.52  &  ...  &  ...  &  ...  &  ...  &  ...  & 5.78  & 7.45  &  ...  &  ...  & 20.91  & 7.29  \\
Fe I & 4840.32 & 4.15 & -1.37 &15.44  & 18.34  &  ...  &  ...  &  ...  &  ...  & 5.59  &  ...  & 18.61  & 7.76  &  ...  &  ...  & 14.73  \\
Fe I & 4843.14 & 3.4 & -1.79 &13.01  &  ...  &  ...  &  ...  &  ...  &  ...  &  ...  &  ...  &  ...  & 8.5  &  ...  &  ...  & 11.55  \\
Fe I & 4844.01 & 3.55 & -2.05 & ...  & 7.08  &  ...  &  ...  &  ...  &  ...  &  ...  &  ...  &  ...  &  ...  &  ...  &  ...  & 9.61  \\
Fe I & 4848.88 & 2.28 & -3.14 &11.0  &  ...  &  ...  &  ...  &  ...  &  ...  &  ...  &  ...  & 8.74  &  ...  &  ...  & 21.66  &  ...  \\
Fe I & 4859.74 & 2.88 & -0.76 & ...  &  ...  & 22.16  &  ...  &  ...  &  ...  &  ...  &  ...  & 72.53  & 70.49  & 34.51  &  ...  & 55.71  \\
Fe I & 4871.32 & 2.87 & -0.36 &94.34  & 97.44  & 34.28  & 43.09  & 44.82  & 62.12  & 72.72  & 103.84  & 110.34  & 100.73  & 58.46  & 134.97  & 83.21  \\
Fe I & 4872.14 & 2.88 & -0.57 &78.32  & 78.33  & 26.47  & 35.35  & 26.95  & 50.2  & 57.8  & 83.18  & 88.31  & 77.35  & 47.04  & 110.34  & 64.51  \\
Fe I & 4875.88 & 3.33 & -1.97 &12.49  & 14.03  &  ...  &  ...  &  ...  &  ...  &  ...  &  ...  & 14.61  & 6.09  &  ...  &  ...  &  ...  \\
Fe I & 4881.72 & 3.3 & -1.78 &14.65  &  ...  &  ...  &  ...  &  ...  &  ...  & 5.37  &  ...  &  ...  &  ...  &  ...  &  ...  & 8.38  \\
Fe I & 4889.0 & 2.2 & -2.55 & ...  &  ...  &  ...  &  ...  &  ...  & 10.15  &  ...  &  ...  &  ...  &  ...  & 5.23  &  ...  &  ...  \\
Fe I & 4891.49 & 2.85 & -0.11 &111.2  & 117.93  & 49.41  & 62.54  & 51.27  & 76.5  & 87.48  & 131.74  & 139.1  & 127.0  & 72.98  & 168.85  & 107.42  \\
Fe I & 4903.31 & 2.88 & -0.93 &72.72  & 75.65  & 17.02  & 25.24  & 17.54  & 38.19  & 48.75  & 73.17  & 79.84  & 71.12  & 33.83  & 105.26  & 61.37  \\
Fe I & 4920.5 & 2.83 & 0.07 &134.03  & 148.25  & 61.51  & 76.55  & 61.6  & 89.51  & 101.31  & 169.58  & 177.64  & 159.56  & 83.6  & 218.67  & 120.04  \\
Fe I & 4927.42 & 3.57 & -2.07 &7.2  &  ...  &  ...  &  ...  &  ...  &  ...  &  ...  &  ...  &  ...  &  ...  &  ...  & 23.43  & 5.78  \\
Fe I & 4938.81 & 2.88 & -1.08 &62.52  & 65.44  &  ...  & 18.69  & 14.07  & 29.05  & 40.53  & 63.04  & 69.52  & 59.05  & 30.85  & 94.06  &  ...  \\
Fe I & 4942.46 & 4.22 & -1.41 &16.6  & 17.7  &  ...  &  ...  &  ...  &  ...  & 6.46  &  ...  & 19.92  & 7.46  &  ...  &  ...  & 8.59  \\
Fe I & 4957.3 & 2.85 & -0.41 & ...  &  ...  & 70.77  &  ...  &  ...  &  ...  &  ...  &  ...  &  ...  &  ...  &  ...  &  ...  &  ...  \\
Fe I & 4962.57 & 4.18 & -1.18 &11.07  & 11.81  &  ...  &  ...  &  ...  &  ...  & 6.7  & 10.57  &  ...  &  ...  &  ...  &  ...  & 7.59  \\
Fe I & 4968.7 & 3.64 & -1.74 &10.96  & 12.1  &  ...  &  ...  &  ...  &  ...  &  ...  &  ...  & 8.51  &  ...  &  ...  & 32.74  & 9.11  \\
Fe I & 4973.1 & 3.96 & -0.92 &30.15  & 32.63  &  ...  & 7.06  &  ...  & 10.04  & 15.94  &  ...  &  ...  & 18.35  & 9.29  & 60.25  & 23.31  \\
Fe I & 4985.25 & 3.93 & -0.56 & ...  & 44.99  &  ...  &  ...  & 6.31  & 17.37  &  ...  &  ...  &  ...  & 40.16  & 14.02  & 67.79  &  ...  \\
Fe I & 4986.22 & 4.22 & -1.37 &7.15  &  ...  &  ...  &  ...  &  ...  &  ...  &  ...  &  ...  &  ...  &  ...  &  ...  &  ...  & 11.72  \\
Fe I & 4991.87 & 4.22 & -1.89 & ...  &  ...  &  ...  &  ...  &  ...  &  ...  &  ...  &  ...  &  ...  &  ...  &  ...  & 8.35  &  ...  \\
Fe I & 4994.13 & 0.91 & -3.08 &60.37  &  ...  & 9.78  & 10.76  & 10.06  & 26.5  &  ...  &  ...  &  ...  & 62.06  & 28.73  & 82.09  &  ...  \\
Fe I & 5002.79 & 3.4 & -1.53 & ...  & 21.66  &  ...  &  ...  &  ...  & 6.34  & 9.5  &  ...  &  ...  & 10.3  & 6.2  &  ...  &  ...  \\
Fe I & 5006.12 & 2.83 & -0.61 &87.01  & 88.24  & 27.47  & 33.97  & 29.46  & 53.98  & 64.69  & 91.09  & 97.35  & 83.33  & 50.79  & 115.72  & 76.32  \\
Fe I & 5021.59 & 4.26 & -0.68 & ...  &  ...  &  ...  &  ...  &  ...  &  ...  &  ...  & 8.2  &  ...  & 5.18  &  ...  & 43.13  & 8.9  \\
Fe I & 5022.79 & 2.99 & -2.2 &23.42  &  ...  &  ...  &  ...  &  ...  & 8.78  &  ...  &  ...  &  ...  &  ...  & 9.0  &  ...  &  ...  \\
Fe I & 5029.62 & 3.41 & -2.0 &7.65  &  ...  &  ...  &  ...  &  ...  &  ...  &  ...  &  ...  &  ...  &  ...  &  ...  &  ...  &  ...  \\
Fe I & 5039.25 & 3.37 & -1.57 &23.65  & 26.86  &  ...  & 7.98  &  ...  & 7.5  &  ...  &  ...  &  ...  & 12.64  & 6.16  & 63.26  & 23.21  \\
Fe I & 5041.76 & 1.48 & -2.2 & ...  &  ...  & 25.8  & 41.25  & 28.64  & 48.25  &  ...  &  ...  &  ...  & 83.97  & 38.49  &  ...  &  ...  \\
Fe I & 5048.44 & 3.96 & -1.03 &13.79  & 15.48  &  ...  &  ...  &  ...  &  ...  &  ...  & 13.31  &  ...  & 7.79  &  ...  &  ...  &  ...  \\
Fe I & 5051.63 & 0.91 & -2.79 &72.0  & 74.84  & 17.85  & 17.83  & 17.22  & 39.41  & 52.08  & 73.18  & 78.21  & 72.27  & 42.05  & 106.32  & 64.5  \\
Fe I & 5056.84 & 4.26 & -1.94 & ...  &  ...  &  ...  &  ...  &  ...  &  ...  &  ...  &  ...  &  ...  & 5.15  &  ...  & 12.15  &  ...  \\
Fe I & 5065.19 & 3.64 & -1.51 & ...  &  ...  &  ...  &  ...  &  ...  & 20.68  &  ...  &  ...  &  ...  &  ...  &  ...  &  ...  &  ...  \\
Fe I & 5074.75 & 4.22 & -0.23 &53.47  & 55.03  & 9.87  & 20.26  & 10.84  & 23.81  & 33.03  & 48.97  & 58.34  & 40.32  & 19.17  & 81.82  & 46.8  \\
Fe I & 5079.74 & 0.99 & -3.22 &47.88  &  ...  & 6.52  & 6.32  & 6.64  &  ...  &  ...  &  ...  &  ...  & 50.52  & 18.33  & 63.97  &  ...  \\
Fe I & 5083.34 & 0.96 & -2.96 &62.52  & 64.78  & 10.88  & 11.57  & 12.78  & 31.33  &  ...  &  ...  & 69.25  & 64.36  & 32.93  & 84.45  &  ...  \\
Fe I & 5090.77 & 4.26 & -0.44 &32.68  & 34.97  &  ...  & 7.24  &  ...  & 11.33  & 15.41  & 26.12  &  ...  & 16.47  & 9.28  & 60.36  & 24.18  \\
Fe I & 5107.45 & 0.99 & -3.09 & ...  &  ...  &  ...  &  ...  &  ...  &  ...  &  ...  &  ...  &  ...  & 53.81  &  ...  &  ...  &  ...  \\
Fe I & 5121.64 & 4.28 & -0.81 &22.61  & 24.59  &  ...  &  ...  &  ...  & 6.33  & 9.69  & 16.65  & 25.87  & 11.77  & 6.74  & 57.18  & 20.66  \\
Fe I & 5126.19 & 4.26 & -1.06 &17.49  & 16.84  &  ...  &  ...  &  ...  &  ...  & 6.14  &  ...  & 19.06  & 8.29  &  ...  & 43.2  &  ...  \\
Fe I & 5131.47 & 2.22 & -2.52 &25.91  &  ...  &  ...  &  ...  &  ...  & 7.71  & 11.98  &  ...  &  ...  & 20.85  & 5.17  & 52.63  &  ...  \\
Fe I & 5137.38 & 4.18 & -0.43 &39.25  & 44.57  & 6.4  & 13.2  & 6.77  & 15.09  & 22.68  &  ...  & 48.43  & 31.01  & 16.43  & 61.45  & 35.93  \\
Fe I & 5139.46 & 2.94 & -0.51 & ...  &  ...  &  ...  & 67.72  &  ...  &  ...  &  ...  &  ...  &  ...  &  ...  &  ...  &  ...  &  ...  \\
Fe I & 5142.93 & 0.96 & -3.07 &57.51  &  ...  & 7.11  &  ...  & 7.23  & 19.08  &  ...  &  ...  &  ...  & 49.85  & 15.72  &  ...  & 44.7  \\
Fe I & 5145.09 & 2.2 & -2.88 &9.18  &  ...  &  ...  &  ...  &  ...  &  ...  &  ...  &  ...  &  ...  & 6.03  &  ...  &  ...  &  ...  \\
Fe I & 5150.84 & 0.99 & -3.04 &56.46  &  ...  & 7.75  & 9.04  & 7.47  & 26.11  &  ...  &  ...  &  ...  & 55.77  & 21.1  & 82.55  &  ...  \\
Fe I & 5191.45 & 3.04 & -0.55 &81.72  & 85.81  & 21.17  & 36.94  & 31.54  & 48.4  & 57.26  & 82.5  & 90.26  & 73.82  & 43.15  & 116.99  & 70.19  \\
Fe I & 5194.94 & 1.56 & -2.09 &74.43  & 74.84  & 19.21  & 24.39  & 24.1  & 44.06  & 54.9  &  ...  & 78.41  & 74.36  & 44.82  & 94.94  & 65.24  \\
Fe I & 5198.71 & 2.22 & -2.13 &46.0  &  ...  & 6.14  & 8.07  & 6.42  & 16.55  & 23.56  &  ...  & 51.54  & 38.6  & 15.54  & 72.51  &  ...  \\
Fe I & 5216.27 & 1.61 & -2.15 &68.63  & 67.94  & 16.14  & 22.4  & 16.43  & 38.2  & 48.02  & 67.49  & 71.81  & 68.32  & 34.06  & 83.55  & 62.54  \\
Fe I & 5217.92 & 3.64 & -1.72 & ...  &  ...  &  ...  &  ...  &  ...  &  ...  &  ...  &  ...  &  ...  &  ...  &  ...  & 23.2  &  ...  \\
Fe I & 5227.19 & 1.56 & -1.23 & ...  &  ...  & 57.91  & 73.33  & 53.74  & 86.02  & 97.46  &  ...  &  ...  & 136.77  & 76.45  &  ...  & 119.4  \\
Fe I & 5232.94 & 2.94 & -0.06 &112.07  & 117.43  & 48.99  & 62.03  & 49.7  & 76.82  & 95.95  & 130.86  & 138.0  & 125.58  & 70.99  & 171.82  & 103.27  \\
Fe I & 5243.78 & 4.26 & -1.12 &14.24  & 13.95  &  ...  &  ...  &  ...  & 5.06  &  ...  &  ...  &  ...  & 6.42  &  ...  &  ...  & 11.3  \\
Fe I & 5249.1 & 4.47 & -1.46 & ...  &  ...  &  ...  &  ...  &  ...  &  ...  &  ...  &  ...  &  ...  &  ...  &  ...  & 12.38  &  ...  \\
Fe I & 5250.65 & 2.2 & -2.18 &49.19  & 50.66  & 6.97  & 9.51  & 6.04  & 20.38  & 25.95  &  ...  &  ...  & 44.32  & 18.22  & 76.89  &  ...  \\
Fe I & 5263.31 & 3.27 & -0.88 &52.22  & 56.17  & 8.11  & 15.25  & 9.03  & 20.55  & 30.08  &  ...  & 62.35  & 44.68  & 24.12  & 83.84  & 43.75  \\
Fe I & 5266.55 & 3.0 & -0.39 &90.93  & 94.44  & 30.74  & 44.01  & 32.28  & 55.72  & 69.06  & 99.47  & 106.86  & 92.62  & 52.92  & 132.83  & 86.05  \\
Fe I & 5269.54 & 0.86 & -1.32 &139.63  & 148.5  & 87.84  & 91.56  & 89.25  & 107.52  & 115.91  & 195.96  & 192.7  & 210.94  & 104.01  & 222.9  & 133.64  \\
Fe I & 5273.16 & 3.29 & -0.99 & ...  &  ...  &  ...  & 19.04  &  ...  &  ...  & 18.42  &  ...  &  ...  & 25.42  &  ...  &  ...  &  ...  \\
Fe I & 5280.36 & 3.64 & -1.82 &6.12  & 6.77  &  ...  &  ...  &  ...  &  ...  &  ...  & 5.08  & 8.41  &  ...  &  ...  & 21.48  &  ...  \\
Fe I & 5283.62 & 3.24 & -0.53 &78.11  & 79.69  & 20.47  & 26.21  & 22.69  & 43.53  & 53.35  & 82.63  & 88.85  & 70.51  & 39.15  & 119.43  & 69.63  \\
Fe I & 5293.96 & 4.14 & -1.84 & ...  &  ...  &  ...  &  ...  &  ...  &  ...  &  ...  &  ...  &  ...  &  ...  &  ...  & 10.79  &  ...  \\
Fe I & 5295.31 & 4.42 & -1.67 & ...  &  ...  &  ...  &  ...  &  ...  &  ...  &  ...  &  ...  &  ...  &  ...  &  ...  & 10.95  &  ...  \\
Fe I & 5302.3 & 3.28 & -0.72 &61.32  & 64.31  & 11.32  & 19.42  & 11.11  & 26.23  & 34.52  & 58.44  & 69.8  & 56.39  & 22.92  & 87.26  & 55.43  \\
Fe I & 5322.04 & 2.28 & -2.8 &11.72  & 13.41  &  ...  &  ...  &  ...  &  ...  &  ...  &  ...  &  ...  & 8.2  &  ...  &  ...  & 8.73  \\
Fe I & 5328.53 & 1.56 & -1.85 & ...  &  ...  & 24.83  & 29.08  & 24.21  & 44.31  & 62.23  & 90.02  & 104.43  &  ...  & 44.05  &  ...  & 75.62  \\
Fe I & 5339.93 & 3.27 & -0.72 &66.34  & 68.81  & 14.1  & 24.22  & 16.01  & 32.78  & 43.2  & 72.66  & 77.94  & 66.11  & 28.12  & 100.25  & 63.71  \\
Fe I & 5361.62 & 4.42 & -1.41 &6.99  &  ...  &  ...  &  ...  &  ...  &  ...  &  ...  &  ...  &  ...  &  ...  &  ...  & 20.61  & 6.28  \\
Fe I & 5364.87 & 4.45 & 0.23 &56.4  & 57.61  & 11.95  & 23.01  & 12.69  & 25.7  & 32.99  & 50.5  & 60.11  & 42.8  & 21.29  & 85.93  & 47.87  \\
Fe I & 5367.47 & 4.42 & 0.44 &61.26  & 65.15  & 14.85  & 30.0  & 15.42  & 36.21  & 40.67  & 57.48  & 69.13  & 51.06  & 27.9  & 93.46  & 52.63  \\
Fe I & 5371.49 & 0.96 & -1.65 &121.69  & 124.19  & 69.85  & 74.24  & 70.63  & 91.45  & 100.52  & 141.71  & 146.97  & 146.45  & 88.47  & 183.75  & 112.16  \\
Fe I & 5387.48 & 4.14 & -2.03 & ...  &  ...  &  ...  &  ...  &  ...  &  ...  &  ...  &  ...  &  ...  &  ...  &  ...  & 10.0  &  ...  \\
Fe I & 5395.22 & 4.45 & -2.15 & ...  &  ...  &  ...  &  ...  &  ...  &  ...  &  ...  &  ...  &  ...  &  ...  &  ...  & 6.25  &  ...  \\
Fe I & 5398.28 & 4.45 & -0.71 &24.96  & 22.7  &  ...  &  ...  &  ...  & 5.81  & 7.84  &  ...  & 22.3  & 8.67  & 5.13  & 41.73  & 15.8  \\
Fe I & 5409.13 & 4.37 & -1.27 &10.16  & 9.9  &  ...  &  ...  &  ...  &  ...  &  ...  & 6.21  & 10.31  &  ...  &  ...  & 26.97  &  ...  \\
Fe I & 5412.78 & 4.43 & -1.72 & ...  &  ...  &  ...  &  ...  &  ...  &  ...  &  ...  &  ...  &  ...  &  ...  &  ...  & 6.41  &  ...  \\
Fe I & 5415.2 & 4.39 & 0.64 &72.72  & 73.96  & 20.42  & 46.14  & 26.22  & 41.52  & 49.0  & 70.69  & 81.4  & 58.4  & 36.36  & 107.92  & 63.53  \\
Fe I & 5429.7 & 0.96 & -1.88 &113.05  &  ...  & 56.91  &  ...  &  ...  &  ...  &  ...  &  ...  & 135.87  &  ...  &  ...  &  ...  & 101.72  \\
Fe I & 6230.72 & 2.56 & -1.28 & ...  & 75.92  &  ...  & 26.94  &  ...  &  ...  &  ...  & 85.09  &  ...  &  ...  & 35.41  & 103.37  &  ...  \\
Fe I & 6232.64 & 3.65 & -1.22 & ...  & 21.81  &  ...  &  ...  &  ...  &  ...  &  ...  & 24.0  &  ...  &  ...  & 5.38  & 53.38  &  ...  \\
Fe I & 6246.32 & 3.6 & -0.88 &47.63  & 49.97  &  ...  & 11.78  & 5.72  & 22.85  & 22.5  & 45.03  & 54.34  & 31.09  & 17.06  & 79.33  & 39.93  \\
Fe I & 6254.26 & 2.28 & -2.43 &41.36  & 46.45  &  ...  & 9.41  & 5.3  & 13.71  & 20.55  & 36.57  & 48.02  & 29.89  & 12.16  & 83.31  & 34.12  \\
Fe I & 6265.13 & 2.18 & -2.55 &29.88  & 32.93  &  ...  &  ...  &  ...  & 12.48  & 14.35  & 31.2  & 38.16  & 26.88  & 8.48  & 61.66  & 24.82  \\
Fe I & 6271.28 & 3.33 & -2.7 & ...  &  ...  &  ...  &  ...  &  ...  &  ...  &  ...  &  ...  &  ...  &  ...  &  ...  & 8.77  &  ...  \\
Fe I & 6297.79 & 2.22 & -2.74 & ...  &  ...  &  ...  &  ...  &  ...  &  ...  & 10.11  &  ...  & 25.8  & 16.49  &  ...  & 44.78  &  ...  \\
Fe I & 6311.5 & 2.83 & -3.14 & ...  &  ...  &  ...  &  ...  &  ...  &  ...  &  ...  &  ...  &  ...  &  ...  &  ...  & 9.16  &  ...  \\
Fe I & 6315.81 & 4.08 & -1.66 & ...  & 5.8  &  ...  &  ...  &  ...  &  ...  &  ...  &  ...  &  ...  &  ...  &  ...  & 19.03  &  ...  \\
Fe I & 6322.68 & 2.59 & -2.43 &22.39  & 24.56  &  ...  &  ...  &  ...  &  ...  &  ...  &  ...  & 21.35  & 14.2  &  ...  & 47.66  & 17.92  \\
Fe I & 6335.33 & 2.2 & -2.18 &42.75  & 45.45  &  ...  & 6.91  & 7.98  & 16.03  & 23.7  & 43.4  & 48.18  & 40.32  & 14.86  & 69.96  & 38.44  \\
Fe I & 6358.63 & 4.14 & -1.66 &13.49  &  ...  &  ...  &  ...  &  ...  &  ...  &  ...  &  ...  &  ...  &  ...  &  ...  &  ...  &  ...  \\
Fe I & 6362.88 & 4.19 & -1.93 & ...  &  ...  &  ...  &  ...  &  ...  &  ...  &  ...  &  ...  &  ...  &  ...  &  ...  & 11.66  &  ...  \\
Fe I & 6380.74 & 4.19 & -1.38 &8.93  & 9.95  &  ...  &  ...  &  ...  &  ...  &  ...  & 7.0  & 10.96  &  ...  &  ...  & 26.4  &  ...  \\
Fe I & 6411.65 & 3.65 & -0.72 &55.59  & 57.66  & 9.44  & 16.09  & 8.62  & 24.7  & 28.94  & 51.68  & 64.2  & 39.1  & 22.03  & 86.66  & 47.79  \\
Fe I & 6421.35 & 2.28 & -2.03 &51.19  & 54.44  & 8.08  & 10.89  & 7.71  & 25.17  & 29.54  & 53.3  & 59.53  & 44.97  & 20.33  & 80.75  & 44.82  \\
Fe I & 6462.71 & 0.91 & -2.17 &111.95  & 118.53  & 35.47  & 61.83  & 51.95  & 68.61  & 74.57  & 107.31  & 127.47  &  ...  & 62.35  & 162.69  & 91.61  \\
Fe I & 6469.19 & 4.83 & -0.81 &8.85  & 10.94  &  ...  &  ...  &  ...  &  ...  &  ...  & 6.32  & 9.63  & 6.75  &  ...  & 24.57  &  ...  \\
Fe I & 6481.87 & 2.28 & -2.98 &11.86  & 17.6  &  ...  &  ...  &  ...  &  ...  & 6.05  & 14.34  &  ...  & 17.01  & 8.1  & 35.66  & 13.54  \\
Fe I & 6495.74 & 4.83 & -0.92 &5.98  & 6.07  &  ...  &  ...  &  ...  &  ...  &  ...  &  ...  &  ...  &  ...  &  ...  & 14.93  &  ...  \\
Fe I & 6533.93 & 4.56 & -1.43 & ...  & 10.2  &  ...  &  ...  &  ...  &  ...  &  ...  &  ...  & 6.64  &  ...  &  ...  & 13.28  &  ...  \\
Fe I & 6569.21 & 4.73 & -0.45 &21.93  & 28.97  &  ...  &  ...  &  ...  & 5.6  & 6.43  & 14.5  & 20.96  & 9.35  &  ...  & 41.69  & 13.92  \\
Fe I & 6575.02 & 2.59 & -2.71 &11.91  & 12.37  &  ...  &  ...  &  ...  &  ...  &  ...  &  ...  & 14.36  & 10.66  &  ...  &  ...  &  ...  \\
Fe I & 6592.91 & 2.73 & -1.47 &52.35  & 55.56  & 7.76  & 10.89  & 8.94  & 26.93  & 30.0  & 53.43  & 63.27  & 44.46  & 21.48  & 80.02  & 42.6  \\
Fe I & 6597.56 & 4.8 & -1.05 &6.26  & 8.17  &  ...  &  ...  &  ...  &  ...  &  ...  &  ...  & 7.38  &  ...  &  ...  & 17.64  &  ...  \\
Fe I & 6752.71 & 4.64 & -1.2 & ...  &  ...  &  ...  &  ...  &  ...  &  ...  &  ...  &  ...  &  ...  &  ...  &  ...  & 13.29  &  ...  \\
Fe I & 6786.86 & 4.19 & -2.02 & ...  &  ...  &  ...  &  ...  &  ...  &  ...  &  ...  &  ...  &  ...  &  ...  &  ...  & 8.9  &  ...  \\
Fe I & 6804.0 & 4.65 & -1.5 & ...  &  ...  &  ...  &  ...  &  ...  &  ...  &  ...  &  ...  &  ...  &  ...  &  ...  & 5.36  &  ...  \\
Fe I & 6820.37 & 4.64 & -1.29 &5.31  & 5.53  &  ...  &  ...  &  ...  &  ...  &  ...  &  ...  & 5.68  &  ...  &  ...  & 16.6  &  ...  \\
Fe I & 6839.83 & 2.56 & -3.35 & ...  &  ...  &  ...  &  ...  &  ...  &  ...  &  ...  &  ...  &  ...  &  ...  &  ...  & 10.08  &  ...  \\
Fe I & 6842.69 & 4.64 & -1.29 &5.72  & 5.67  &  ...  &  ...  &  ...  &  ...  &  ...  &  ...  &  ...  &  ...  &  ...  & 15.29  &  ...  \\
Fe I & 6858.15 & 4.61 & -0.93 &8.97  & 8.84  &  ...  &  ...  &  ...  &  ...  &  ...  & 6.43  & 9.2  &  ...  &  ...  & 23.25  &  ...  \\
Fe I & 6885.75 & 4.65 & -1.35 & ...  &  ...  &  ...  &  ...  &  ...  &  ...  &  ...  &  ...  &  ...  &  ...  &  ...  & 13.56  &  ...  \\
Fe I & 6945.2 & 2.42 & -2.48 &23.27  & 26.11  &  ...  &  ...  &  ...  &  ...  &  ...  & 23.58  & 48.61  & 38.05  &  ...  & 54.17  & 22.53  \\
Fe I & 6978.85 & 2.48 & -2.5 &29.39  & 22.71  &  ...  &  ...  &  ...  & 7.35  & 13.13  & 22.02  & 26.6  & 18.43  &  ...  & 51.25  & 23.38  \\
Fe I & 7016.39 & 4.15 & -1.21 &10.17  & 19.33  &  ...  &  ...  &  ...  &  ...  &  ...  & 7.94  &  ...  &  ...  &  ...  & 24.14  &  ...  \\
Fe I & 7038.22 & 4.22 & -1.25 &12.83  & 12.54  &  ...  &  ...  &  ...  &  ...  &  ...  &  ...  & 13.17  &  ...  &  ...  & 31.47  &  ...  \\
Fe I & 7090.38 & 4.23 & -1.16 &13.03  & 13.0  &  ...  &  ...  &  ...  & 6.81  &  ...  & 9.89  & 15.77  & 5.69  &  ...  & 31.6  &  ...  \\
Fe I & 7132.99 & 4.08 & -1.63 & ...  & 7.51  &  ...  &  ...  &  ...  &  ...  &  ...  &  ...  & 7.9  & 7.02  &  ...  & 18.99  &  ...  \\
Fe I & 7223.66 & 3.02 & -2.21 &13.73  & 14.1  &  ...  &  ...  &  ...  &  ...  &  ...  & 13.08  &  ...  &  ...  &  ...  & 36.3  &  ...  \\
Fe I & 7320.68 & 4.56 & -1.16 & ...  &  ...  &  ...  &  ...  &  ...  &  ...  &  ...  & 5.39  &  ...  &  ...  &  ...  &  ...  &  ...  \\
Fe I & 7401.68 & 4.19 & -1.35 &5.86  & 5.97  &  ...  &  ...  &  ...  &  ...  &  ...  & 5.99  & 6.36  & 6.72  &  ...  & 18.08  &  ...  \\
Fe II & 4923.92 & 2.89 & -1.21 & ...  &  ...  & 61.07  & 98.8  & 68.36  & 85.16  & 87.14  &  ...  &  ...  & 67.26  & 74.75  &  ...  & 87.54  \\
Fe II & 4993.35 & 2.81 & -3.7 &14.69  &  ...  &  ...  &  ...  &  ...  &  ...  & 6.15  &  ...  &  ...  & 5.18  &  ...  &  ...  &  ...  \\
Fe II & 5018.44 & 2.89 & -1.35 &128.21  & 123.3  & 70.3  &  ...  & 78.67  & 94.38  & 93.62  & 88.48  & 110.97  & 70.94  & 91.68  & 174.78  & 104.18  \\
Fe II & 5169.03 & 2.89 & -0.87 &171.52  & 170.11  & 84.76  & 118.12  & 92.37  & 128.49  & 127.63  & 148.12  & 166.03  & 123.73  & 118.06  & 224.53  & 144.14  \\
Fe II & 5197.57 & 3.23 & -2.05 &56.14  & 52.55  & 13.19  &  ...  & 16.92  & 32.42  & 34.12  &  ...  &  ...  & 14.72  & 23.82  & 64.5  & 42.88  \\
Fe II & 5234.62 & 3.22 & -2.21 &62.39  & 58.04  & 15.36  & 39.93  & 20.25  & 35.53  & 38.73  &  ...  &  ...  & 19.49  & 26.8  & 67.6  & 52.91  \\
Fe II & 5264.8 & 3.23 & -3.23 &18.58  & 23.48  &  ...  & 7.75  &  ...  & 7.05  &  ...  &  ...  &  ...  &  ...  &  ...  &  ...  &  ...  \\
Fe II & 5276.0 & 3.2 & -1.9 & ...  &  ...  & 17.82  & 46.79  & 21.28  & 39.31  & 41.56  & 35.51  &  ...  & 25.65  & 30.82  &  ...  & 49.99  \\
Fe II & 5284.1 & 2.89 & -3.2 &29.55  &  ...  &  ...  & 10.24  &  ...  & 11.65  & 13.07  & 11.24  &  ...  & 5.97  &  ...  &  ...  & 19.82  \\
Fe II & 5325.55 & 3.22 & -3.26 &21.1  & 14.11  &  ...  & 6.55  &  ...  & 6.09  &  ...  & 5.33  &  ...  &  ...  &  ...  &  ...  &  ...  \\
Fe II & 5414.07 & 3.22 & -3.48 &9.52  &  ...  &  ...  &  ...  &  ...  &  ...  &  ...  &  ...  &  ...  &  ...  &  ...  & 13.45  & 7.44  \\
Fe II & 5425.25 & 3.2 & -3.4 &14.01  &  ...  &  ...  &  ...  &  ...  & 5.85  & 5.72  &  ...  &  ...  &  ...  &  ...  &  ...  & 16.31  \\
Fe II & 6238.39 & 3.89 & -2.8 &16.23  & 15.9  &  ...  & 8.26  &  ...  & 7.01  & 5.82  & 5.35  & 9.93  &  ...  &  ...  & 27.33  &  ...  \\
Fe II & 6247.56 & 3.89 & -2.4 &27.82  & 23.2  &  ...  & 13.69  & 5.7  & 14.11  & 12.21  & 9.97  & 17.34  &  ...  & 6.38  & 34.77  & 17.85  \\
Fe II & 6416.92 & 3.89 & -2.9 &15.78  & 12.79  &  ...  & 6.44  &  ...  & 5.89  & 6.11  & 6.04  & 8.18  &  ...  &  ...  & 21.65  &  ...  \\
Fe II & 6432.68 & 2.89 & -3.5 &15.87  & 13.75  &  ...  & 6.64  &  ...  & 5.44  & 6.33  & 6.27  & 10.64  &  ...  &  ...  & 21.88  &  ...  \\
Fe II & 6456.38 & 3.9 & -2.2 &37.7  & 27.7  & 6.36  & 14.96  & 8.33  & 18.17  & 21.76  & 14.16  & 24.18  & 6.24  & 11.05  & 44.9  & 26.26  \\
Fe II & 6516.08 & 2.89 & -3.37 &21.96  & 19.68  &  ...  & 16.21  &  ...  & 8.53  & 10.31  & 10.51  & 16.72  &  ...  & 17.13  & 58.95  & 16.44  \\
Fe II & 7222.39 & 3.89 & -3.4 & ...  &  ...  &  ...  &  ...  &  ...  &  ...  &  ...  &  ...  &  ...  &  ...  &  ...  & 14.85  &  ...  \\
Fe II & 7224.48 & 3.89 & -3.4 & ...  &  ...  &  ...  &  ...  &  ...  &  ...  &  ...  &  ...  &  ...  &  ...  &  ...  & 11.17  &  ...  \\
\enddata
\tablecomments{Only those lines that we were able to measure for at least one star are listed. The line information section includes element, wavelength, excitation potential, and line strength data. Lines measured as weaker than the 5m\AA\ cutoff are not included in this table.}
\end{deluxetable}
\end{longrotatetable}

\end{CJK*}
\end{document}